\documentclass{aa}
\usepackage{newtxtext,newtxmath}
\usepackage{txfonts}
\usepackage{hyperref}

\usepackage[T1]{fontenc}

\DeclareRobustCommand{\VAN}[3]{#2}
\let\VANthebibliography\thebibliography
\def\thebibliography{\DeclareRobustCommand{\VAN}[3]{##3}\VANthebibliography}

\usepackage{xr}
\usepackage{xcolor}
\usepackage{graphicx}	
\usepackage{amsmath}	

\usepackage{verbatim}
\usepackage{breqn}
\usepackage[export]{adjustbox}
\usepackage{booktabs}
\usepackage{pgfplotstable}
\usepackage{longtable}
 \usepackage{threeparttable}
\usepackage{pgfplots} 
\pgfplotsset{compat=1.18}
\usepackage{footnote}
\usepackage{array}
\makesavenoteenv{tabular}

\newcommand{\Ndot}{\overset{\displaystyle .}{N}}

\newcommand{\lya}{Ly$\alpha$}

\newcommand{\kms}{km~s$^{-1}$}
\newcommand{\hii}{\mbox{\tiny H\,{\sc ii}}}
\newcommand{\hi}{\mbox{\tiny H\,{\sc i}}}
\newcommand{\HI}{\mbox{H\,{\sc i}}}
\newcommand{\HII}{\mbox{H\,{\sc ii}}}

\newcommand{\HeI}{\mbox{He\,{\sc i}}}
\newcommand{\HeII}{\mbox{He\,{\sc ii}}}

\newcommand{\NHI} {$N_{\rm HI}$}
\newcommand{\xhi}{$x_{\mathrm{HI}}$}

\begin{document}
\title{On the Origin of the Ly$\alpha$ Damping Wing in Galaxies at $8\le z \le 10$: Explorations using the NINJA Simulations}
\titlerunning{\lya\ absorption from high-$z$ galaxies}

%

 \author{Sukanya Mallik
          \inst{1}\fnmsep\thanks{E-mail: sukanya.mallik@inaf.it}
          \and
          Raghunathan Srianand\inst{2}
          \and 
          Nishikanta Khandai\inst{3,4}
          }

   \institute{Istituto Nazionale di Astrofisica – Osservatorio Astronomico di Trieste, Via Tiepolo 11, I-34143 Trieste, Italy
         \and   
             IUCAA, Postbag 4, Ganeshkhind, Pune 411007, India
         \and  
             School of Physical Sciences, National Institute of Science Education and Research, Jatni, Odisha 752050, India 
         \and
            Homi Bhabha National Institute, Training School Complex, Anushaktinagar, Mumbai 400094, India
             }  
\date{Accepted XXX. Received YYY; in original form ZZZ}

\abstract
   {}
   {
   Ly$\alpha$ damping wing measurements of galaxies at $7 \leq z \leq 14$ with the {\it James Webb Space Telescope} (JWST) provide a powerful probe of the Epoch of Reionization. We combine the {\sc Ninja} hydrodynamical simulations with an idealized ionized-bubble model at $z=8$ and $10$ to quantify the contributions of the intergalactic medium (IGM), circumgalactic medium (CGM), and interstellar medium (ISM) to the observed Ly$\alpha$ damping wing.
   } 
  %
 {
  We use large-scale cosmological hydrodynamical simulations including star formation, stellar feedback, and a uniform ionizing background to generate mock Ly$\alpha$ absorption spectra at $z=8$ and 10. We consider three idealized scenarios: (i) galaxies embedded in a uniformly ionized IGM, (ii) galaxies surrounded by H~\textsc{ii} regions with radii of 0–400 pkpc, and (iii) the same models including Ly$\alpha$ absorption from partially ionized gas within the virial radius. We explore IGM neutral fractions of $x_{\mathrm{HI}}=0.1$, 0.8, and 1.0, and compare the distribution of inferred H~\textsc{i} column densities with observations to constrain the sizes of H~\textsc{ii} regions and the ionization state of the IGM and gas in and around galaxies.
  }
   {
We find that galaxies are surrounded by over-dense gas extending to $\sim60-$100 pkpc, with its extent increasing with halo mass and showing little evolution between $z=8$ and 10. The inferred H~\textsc{i} column density also increases with halo and stellar mass. While the observed incidence of damped Ly$\alpha$ absorption and the median H~\textsc{i} column density can be reproduced by different combinations of the IGM neutral fraction and H~\textsc{ii} region size, models with a uniformly ionized IGM or H~\textsc{ii} regions larger than $\sim50$~pkpc fail to reproduce the strongest absorbers ($N_{\mathrm{H~\textsc{i}}}>10^{22}~\mathrm{cm^{-2}}$). These absorbers preferentially arise in massive halos when absorption from partially ionized gas within the virial radius is included, although their predicted incidence remains below the observations, suggesting an additional contribution from unresolved gas in stellar birth clouds. The strongest damped Ly$\alpha$ absorbers therefore provide a unique probe of the ionization state of the ISM, CGM, and IGM.
}
   {}

 




   

\maketitle


\section{Introduction}
\label{Sec:introduction}
The physical state and chemical enrichment of the intergalactic medium (IGM) are intimately connected to the formation of the first stars and galaxies and the associated feedback processes. Ultraviolet (UV) photons from these early sources ionized the surrounding \HI\ and \HeI\ into H~{\sc ii} and \HeII, respectively.  These ionized regions, or bubbles, gradually expanded and eventually percolated throughout the Universe, marking the completion of the "Epoch of Reionization (EoR)". Observations indicate that hydrogen reionization was largely complete by $z\sim6$ \citep{Barkana2001, Choudhury2001, Ciardi2005, Furlanetto2006, Loeb2001, Morales2010, Zaroubi2013}. The progress of reionization is commonly described by the volume-averaged neutral hydrogen fraction, \xhi. However, even a small \xhi\ can saturate \lya\ absorption in the spectra of luminous background sources near the EoR, making \xhi\ difficult to measure directly from \lya\ absorption alone. Consequently, complementary probes, including \lya\ emission-line equivalent width distribution \citep{mason2018, mason2019, bolan2022, morishita2023, Jones2024}, luminosity function  \citep{Inoue2018, Morales2021} and clustering of \lya\ emitters \citep{sobacchi2015, ouchi2018}, and the \lya\ forest \citep{McGreer2015, Jin2023}, are widely used to constrain the ionization state of the IGM.

The \lya\ damping wing observed redward of the \lya\ emission line at the source redshift, has long been used to constrain the neutral hydrogen fraction in the diffuse IGM using quasar (QSO) spectra \citep[e.g.][]{Fan2006, Bolton2011, grieg2017, Davies2018a, Grieg2019}. QSOs are well suited for such studies because they are sufficiently bright to be observed with ground-based telescopes \citep[see, e.g.,][]{Dodorico2023} and their intense ionizing radiation with near unity escape fraction of hydrogen-ionizing photons, ionize most of the local \HI\ within the proximity zone \citep[e.g.][]{cen2000}. However, the inferred damping wing depends on both the IGM neutral fraction and the poorly constrained QSO lifetime, introducing a significant degeneracy \citet{Hennawi2025}.

The advent of the \textit{James Webb Space Telescope} (JWST) has extended \lya\ damping-wing studies to much fainter, but far more numerous, galaxies at $7\le z\le14$, providing a powerful new probe of the EoR \citep[e.g.][]{Curtis-Lake2023, Heintz2024, Hsiao2023}. Using  an MCMC analysis of stacked JWST/NIRSpec spectra of 27 galaxies, \citet{Umeda2023} simultaneously constrained the IGM neutral fraction and the sizes of the surrounding ionized bubbles, finding \xhi\ to increase from 0.53 to 0.92 over $z=7.12-9.91$. More recently, \citet{Umeda2026} inferred \xhi\ $\approx 0.65$ at $z\sim7$ and an almost fully neutral IGM at $z\ge9$ from a substantially larger galaxy sample observed through multiple JWST/NIRSpec spectroscopy programs.

Despite these exciting developments, interpreting Ly\(\alpha\) damping wings in galaxy spectra remains considerably more challenging than in brighter sources. Since high-redshift galaxies are intrinsically faint, most observations are obtained with the highly sensitive JWST/NIRSpec PRISM mode, which provides only modest spectral resolution (\(R\sim100\)--300). Because of this low spectral resolution, the Ly\(\alpha\) damping wing is sampled by only a few spectral resolution elements. Consequently, the characteristic shape of the damping wing is encoded in only a small number of data points, leading to strong degeneracies between absorption arising from the IGM, the circumgalactic medium (CGM), and the galaxy's interstellar medium (ISM), thereby limiting the robustness of inferred IGM neutral fractions. The interpretation is further complicated by the intrinsic diversity of star-forming galaxy spectra, arising from variations in stellar populations, dust attenuation, nebular continuum emission, and emission-line strengths, all of which affect the intrinsic Ly\(\alpha\) profile.

A key question,  therefore, is about the relative contribution of the different neutral gas components to the observed \lya\ damping wing. While the ionization state of the IGM is of primary interest, significant absorption can also arise from neutral gas associated with the galaxy itself, particularly when the H~{\sc i} column density reaches the damped Ly\(\alpha\) regime (\(N_{\mathrm{HI}}\gtrsim10^{21}~\mathrm{cm}^{-2}\)). Several JWST/NIRSpec studies have indeed reported such high column densities in high-$z$ galaxies \citep[e.g.][]{Umeda2023, Hsiao2023, Curtis-Lake2023, Heintz2024}. \citet{Heintz2025} found that \(65\)--\(90\%\) of galaxies in the JWST-PRIMAL sample exhibit \(N_{\mathrm{HI}}>10^{21}~\mathrm{cm}^{-2}\) at \(z\ge8\), suggesting that galaxy-associated gas makes a substantial contribution to the observed \lya\ damping wing. Likewise, \citet{Pollock2026} reported that the majority of galaxies at \(z>9\) show prominent DLA signatures with  \(N_{\mathrm{HI}}\gtrsim10^{22.5}~\mathrm{cm}^{-2}\), while \citet{Mason2025} found that \(18\%\) of their JWST/NIRSpec prism sample spanning \(z=5.5\)--13 has \(N_{\mathrm{HI}}\gtrsim10^{22}~\mathrm{cm}^{-2}\). By comparing these observations with the semi-numerical reionization simulations of \citet{Lu2024}, \citet{Mason2025} further inferred that the highest-column-density absorbers contributing to the \lya\ damping wing predominantly arise within \(\lesssim500\) pkpc of the source galaxy. All these demonstrate that the observed \lya\ damping wing is generally a composite of absorption from the IGM and galaxy-associated gas. Consequently, robust constraints on the \xhi\ of the IGM  require physically motivated models that simultaneously account for absorption by the IGM, the circumgalactic environment, and the ISM.

On the simulation side, significant efforts have been made to capture the physics of reionization across different spatial scales. Cosmological simulations incorporating radiative transport (RT), such as the Sherwood-Relic simulations \citep{Puchwein2023}, have been developed to model the large-scale ionization structure of the Universe during reionization. In parallel, high-resolution zoom-in simulations with RT have been used to investigate the local contributions to Ly\(\alpha\) absorption from star-forming regions and the circumgalactic medium (CGM) of galaxies \citep[e.g.][]{Kannan2025,Gelli2025,Steen2026}.

Using the Sherwood-Relic simulations, \citet{Keating2024} demonstrated that the average neutral hydrogen density along the line of sight determines the characteristic shape and scatter of the Ly\(\alpha\) damping wing. They found median ionized bubble sizes of 11.7 and 6.4 cMpc at \(z\sim7\) and 8, respectively. Using the same simulations, \citet{Keating2023} showed that simulated Ly\(\alpha\) damping wing profiles around halos exhibit stronger absorption than expected from neutral hydrogen in the diffuse IGM alone, suggesting an important contribution from gas associated with galaxies.
On the other hand, zoom-in simulations have been particularly valuable for quantifying the contribution of the ISM and CGM of galaxies to the observed Ly\(\alpha\) damping wing. These simulations also provide insights into possible correlations between line-of-sight H~{\sc i} column densities and the physical properties of galaxies. Together, large-scale cosmological simulations and high-resolution zoom-in simulations provide complementary views of the different physical components contributing to the observed Ly\(\alpha\) damping wing.

In this work, we use cosmological hydrodynamical simulations from the {\sc Ninja} 
\citep[{\bf N}ISER\footnote{\href{https://www.niser.ac.in}{https://www.niser.ac.in/}}-
{\bf I}UCAA\footnote{\href{https://www.iucaa.in/en/}~{https://www.iucaa.in/en/}} 
{\bf N}ew Simulations of 
{\bf J}WST
G{\bf A}laxies and Quasars, as described in detail in][]{Behera2026}
simulation suite, to investigate the contributions to the \lya\ damping wing from (i) \HI\ in the uniformly ionized diffuse IGM, (ii) gas clustered around star-forming galaxies, and (iii) \HI\ gas intrinsic to the galaxies. To this end, we employ an idealized model in which halos are surrounded by ionized bubbles embedded in a neutral (or partially ionized) IGM. By varying the ionized bubble size and the global IGM neutral fraction, we explore a range of reionization scenarios. Although our simulations do not include self-consistent RT, the adopted models capture the key physical environments expected during the EoR.

This paper is organized as follows. Section~2 describes the simulations used and Section~3 presents the halo identification procedure, the generation of mock absorption spectra along sightlines towards halos and through the IGM, and the parameters used to characterize the \lya\ absorption. In Section~4, we present the radial overdensity profiles and \lya\ damping wing profiles for three models that separately include contributions from the IGM, the IGM and \HII\ regions around galaxies, and the IGM, \HII\ regions, and residual neutral gas within galaxies. Section~5 discusses the convergence of our results and their implications. Finally, Results and conclusions are presented in Section~6.

\section{Details of simulations:}
\label{sec:simu}

\begin{figure*}
\centering
\includegraphics[width=0.8\textwidth]{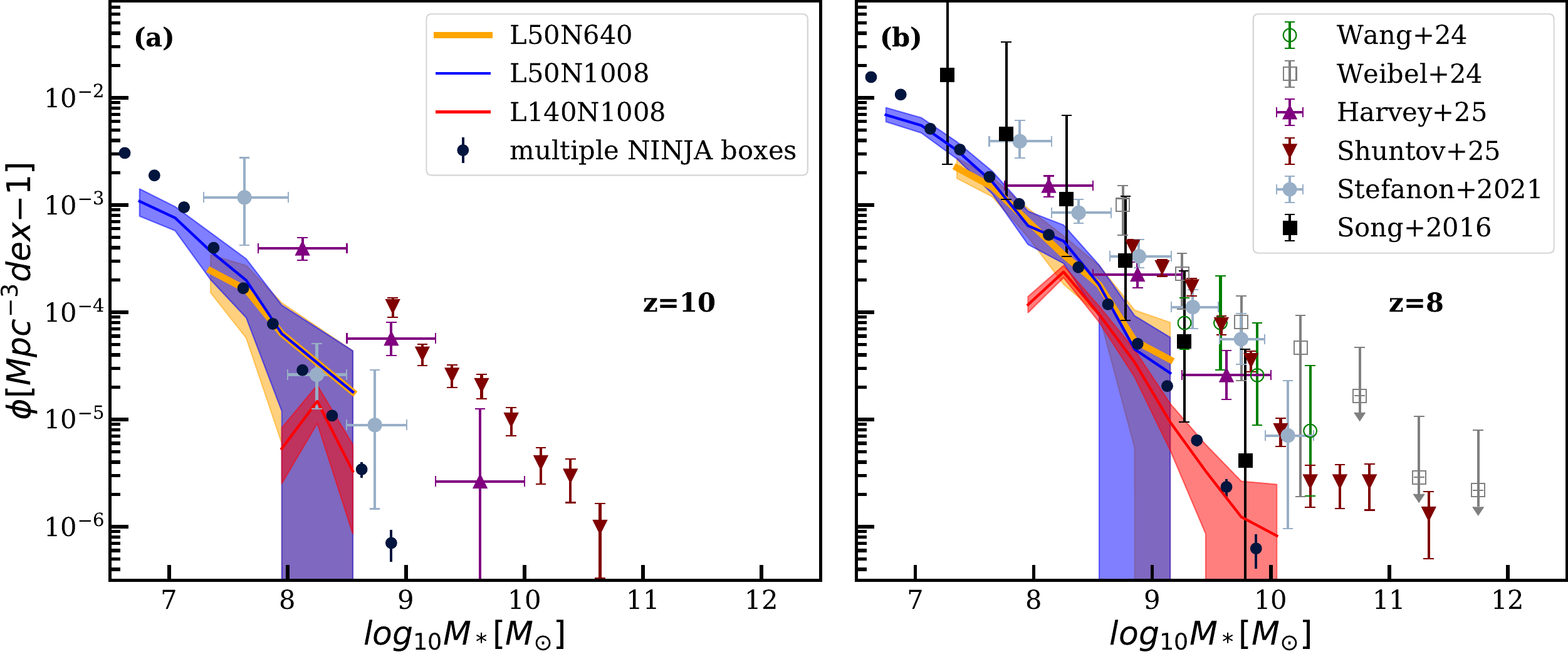}
	\caption{The galaxy stellar mass function at $z = 10,8$ are shown in the left and right panels
    for the three simulations used in our work, i.e., L50N640, L50N1008 and L140N1008. The stellar mass function of the halos in the simulation snapshot is obtained using the total stellar mass of each halo, identified using halo finding software {\sc Rockstar Galaxies}. The combined galaxy stellar mass function obtained from the FOF selected halos of multiple high resolution {\sc Ninja} boxes from \citet{Behera2026} is shown in black circles. Our low resolution simulations follow the general trend.
    The observation data collected from \citet{song2016}, \citet{stefanon2021}, \citet{harvey2025}, and \citet{shuntov2025} are shown in square and circular markers. 
    Our simulations seem to under-produce the GSMF compared to the observations shown here. 
    }
\label{fig:GSMF}
\end{figure*}
We use three boxes from the {\sc Ninja} simulation suite (See Table~\ref{tab:sim_details}), developed to study the physics of galaxies at high-$z$. The simulations are performed with {\sc MP-Gadget} \citep{feng2018}, a massively scalable cosmological hydrodynamical code based on {\sc P-Gadget-3} \citep{Springel2005,dimatteo2012,Khandai2015}. The adopted subgrid models for star formation, black hole growth, and feedback closely follow those implemented in {\sc Illustris-TNG} \citep{weinberger2017,pillepich2018} and {\sc Astrid} \citep{bird2022}. The simulation boxes in this paper \citep[see also,][]{2025arXiv251114739M} represent the fiducial runs, while higher-resolution {\sc Ninja} simulations are described in \citet{Behera2026}. We adopt the WMAP-9 cosmological parameters,  $\left( \Omega_m, \Omega_b,\Omega_{\Lambda}, \sigma_8, n_s, h\right)=\left\{0.2814, 0.0464, 0.7186, 0.81, 0.971, 0.697\right)$ \citep{Hinshaw2013}. Initial conditions are generated at z=99 using CLASS \citep{Lesgourgues2011}.

The star formation model follows \citet{feng2016}, based on the multiphase model of \citet{springel2003a}, and incorporates cooling from primordial gas \citep{katz1996} and metal-line cooling \citep{vogelsberger2013}, self-shielding corrections for neutral hydrogen in dense regions \citep{rahmati2013}, and molecular-hydrogen-regulated star formation \citep{2011ApJ...729...36K}. Stellar evolution, chemical enrichment from massive stars and supernovae are implemented following \citep{vogelsberger2013,pillepich2018}, assuming a \citet{chabrier2003} initial mass function(IMF). A variable stellar wind feedback which depends on the local dark matter velocity dispersion is implemented as in \cite{okamoto2010}.

The supermassive black hole model follows \citet{dimatteo2005, springel2005a,2008ApJ...676...33D}. Black holes are seeded with a constant seed mass of $5 \times 10^5$ $h^{-1}$ M$_{\odot}$ in newly formed friends-of-friends \citep[FOF,][]{1985ApJ...292..371D}
halos with a minimum total mass  $M^{\rm FOF}=10^{10}$ $h^{-1}$M$_{\odot}$ and a minimum stellar mass $M^{\rm FOF}_{\star} =5\times 10^{6}$ $h^{-1}$M$_{\odot}$ if it does not already contain one.
Black holes accrete at the Bondi-Hoyle-Littleton \citep{1939PCPS...35..405H,1944MNRAS.104..273B} accretion rate with boost factor, $\alpha = 100$,  to account for the unresolved ISM.
The accretion rate is limited by an upper cap of $2 \times$ Eddington rate. Black holes experience dynamical friction \citep{1943ApJ....97..255C} due to collisionless particles, forcing them 
to remain close to the center of the halo \citep{2022MNRAS.510..531C}. Black hole mergers are treated following the {\sc Astrid} implementation \citep{2022MNRAS.513..670N}. Feedback switches between thermal-mode (high accretion rate) and kinetic-mode (low accretion rate) similar to {\sc Illustris-Tng} \citep{weinberger2017}. However, since kinetic mode typically sets in at $z\le2$, whereas the thermal mode is the dominant mode at higher redshifts, kinetic mode has little effect on the high-redshift galaxies analysed here. Black holes radiate with a bolometric luminosity $L_{\rm bol}=\eta \dot{M}_{\rm bh}c^2$, where $\eta=0.1$ is the radiative efficiency \citep{1973A&A....24..337S}. In the thermal feedback mode, 5\% of the radiated energy couples to the surrounding gas.

\begin{figure*}
\centering
\includegraphics[width=0.8\textwidth, trim=0 8cm 0 0, clip]{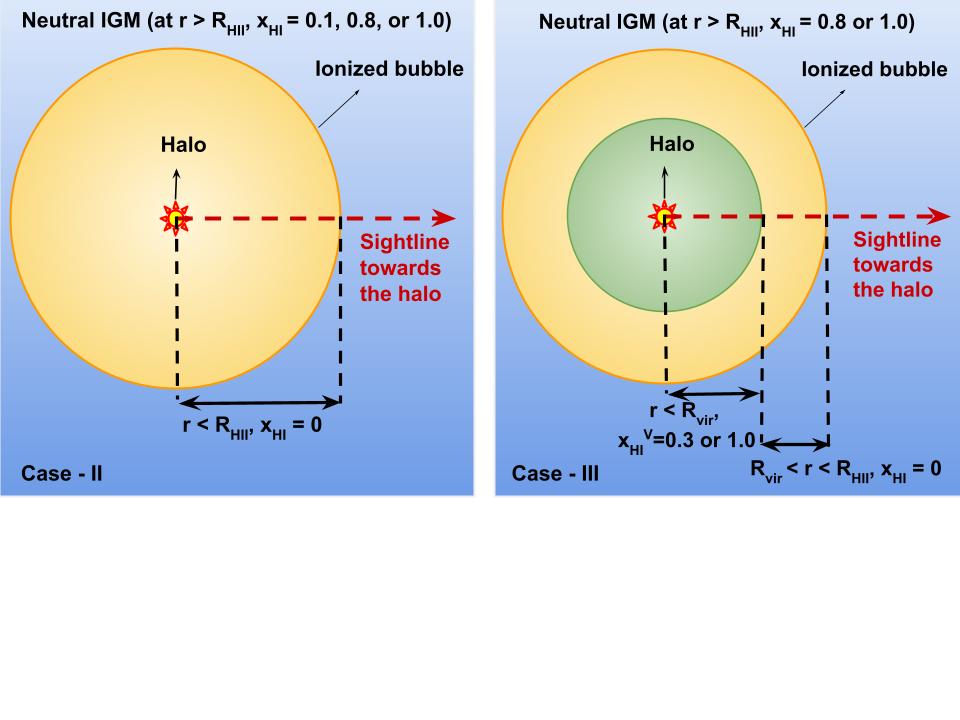}
	\caption{The schematic diagram of the toy-models used in this work. The left panel shows the model for Case-II, where each halo is assumed to be surrounded by an ionized bubble of radius $R_{\rm HII}$, with no neutral gas inside this bubble. The medium outside this bubble has neutral fraction (\xhi). We vary the values of $R_{\rm HII}$ from 0-400~pkpc. The value of \xhi is varied among 0.1, 8.8, and 1.0. In the right panel we show the model corresponding to Case-III, which includes the contribution from the residual H~{\sc i} within the virial radius($R_{\mathrm{vir}}$) of the halos. The neutral hydrogen fractions within $R_{\mathrm{vir}}$ is $x_{\mathrm{HI}}^{\rm V}$ in this modified toy-model. We discussed the results for two scenarios with $x_{\mathrm{HI}}^{\rm V}$ values of 0.3 or 1.0 and \xhi of the IGM of 0.8 or 1.0.}
\label{fig:model1}
\end{figure*}

Our simulations were run using a spatially uniform but time-varying UV-background (UVB) given by \citet[][December 2011 update of the spectrum]{faucher2009}. For the redshift range probed here we expect the ionization state of the IGM to be patchy and this needs to be taken into account while studying the \lya\ damping wing in the galaxy spectra. 
Several hydrodynamical simulations consider radiative transfer (RT) either as a post-processing step [e.g., using RT codes on the outputs of {\sc MassiveBlack-II} simulation \citep{eide2018, eide2020} or on the outputs of {\sc Illustris} \citep{bauer2015}] or using approximate RT algorithm without using a numerical solver [as done in {\sc BlueTides} \citep{feng2016}, {\sc Astrid} \citep{bird2022}] or fully coupled with the simulation run [e.g., {\sc Croc} \citep{gnedin2014}, {\sc CosmicDawn} \citep{Ocvirk2016, Ocvirk2020}, {\sc TechnicolorDawn} \citep{Finlator2018}, {\sc Sphinx} \citep{Rosdahl2018},  {\sc Thesan} \citep{kannan2022}, etc.]. Since there are always uncertainties related to the photon production rate, escape fraction, dust content, etc., here we approach the problem using idealized models (shown in Figure~\ref{fig:model1}) where we consider highly ionized regions (with the flexibility to vary its properties) around halos and neutral or partially ionized gas  in the rest of the IGM.  The details of this model, generation of mock spectra and characterization of \lya\ damping wing profile are presented below.

\begin{figure*}
\begin{minipage}{\textwidth}
\centering
\includegraphics[width=0.7\textwidth]{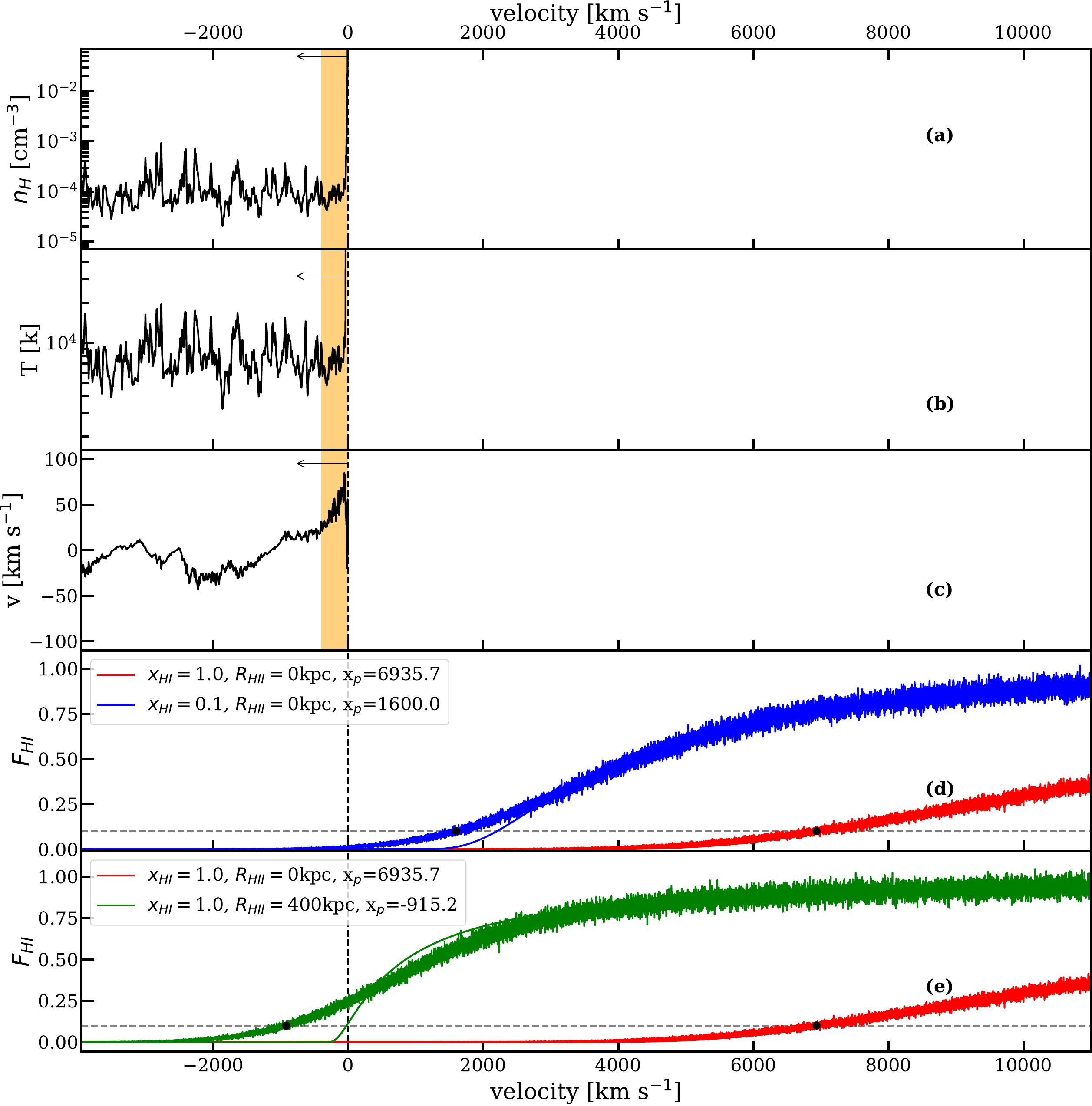}
	\caption{Example of a spectrum for case~II: Panels (a), (b), and (c) show the hydrogen number density $n_H$, temperature ($T$), peculiar velocity ($v$) along a sightline drawn from the center of mass of the stars of a halo, situated at the redshift indicated by the black dashed line in all the panels. An arrow in these panels shows the direction of the light along the sightline. The \lya\ profiles, corresponding to the two different IGM neutral fractions (  $x_{\mathrm{HI}} =0.1 $ and $1.0$), are shown in panel (d). The \lya\ profiles, before considering the effects of instrumental resolution and noise, are in fainter colors. The bright blue and red lines show the profiles obtained after considering an instrumental resolution of R$=100$ (appropriate for PRISM at $\lambda < 2\mu m$) and noise corresponding to SNR 30. The yellow shaded region in panels (a), (b), and (c) marks the ionized region of 400~pkpc around the galaxy. Panel (e) shows the \lya\ profile for the ionized bubble radius ($R_{\rm HII}$) of 0 and 400~pkpc in red and green lines. The strength of the absorption is parameterized by $x_p$, defined as the velocity separation between the source and the $10\%$ transmission (indicated by the black dots). The $x_p$ values are mentioned for different $x_{\mathrm{HI}}$ and $R_{\rm HII}$ in panels (d) and (e). 
	}
\label{fig:demo_los}
\end{minipage}%
\end{figure*}

\begin{figure}
 	\begin{center}
	\includegraphics[width=0.47\textwidth]{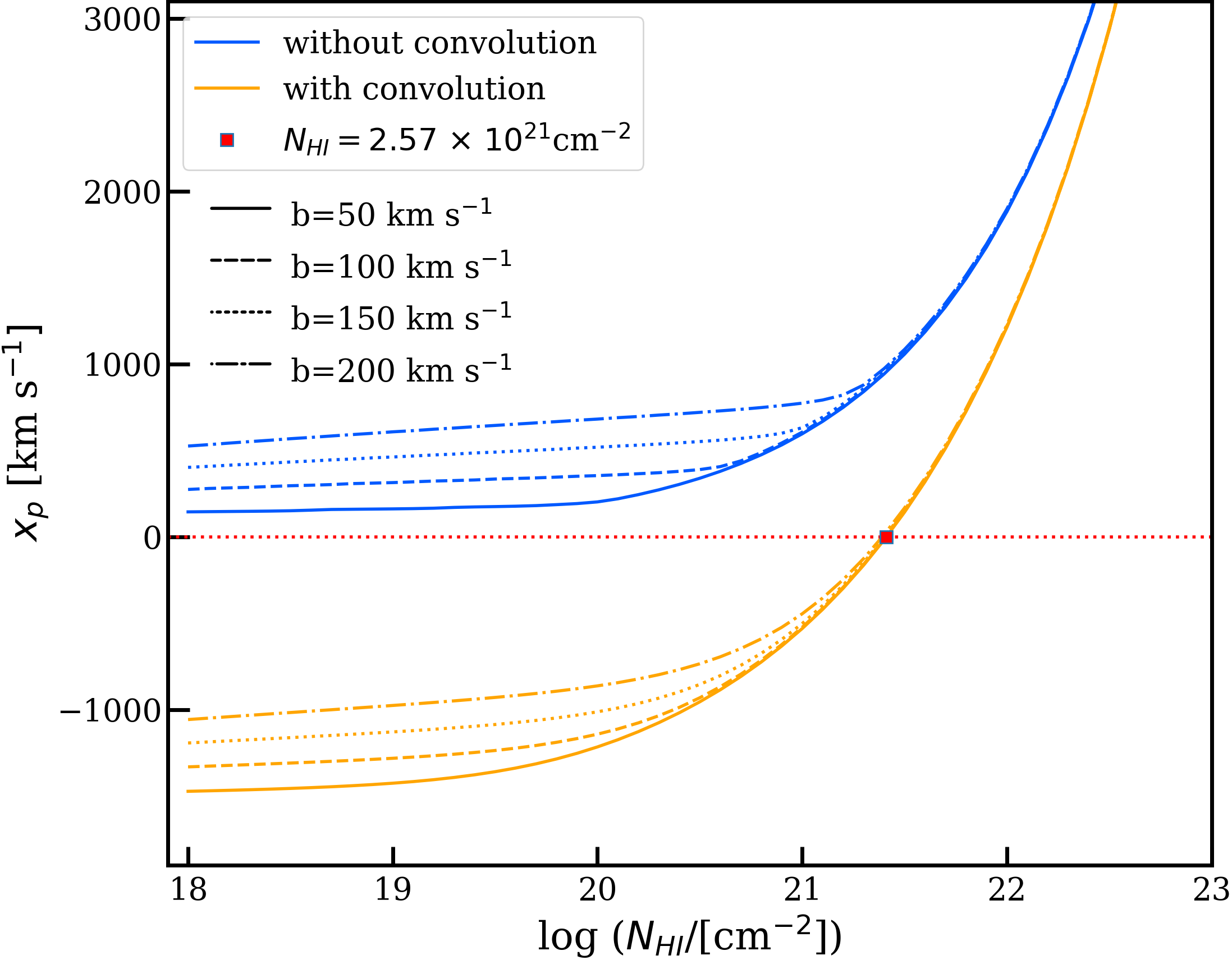}
		\end{center}
\caption{The mapping between $x_p$ and \NHI\ for a single DLA. Results before and after including the effect of instrumental broadening are shown in blue and orange respectively. Evidently, the $x_p$ values are highly affected by the instrumental broadening and the effect of internal velocity dispersions are insignificant for \NHI $>10^{21} ~\mathrm{cm}^{-2}$. We estimate \NHI\ from the measured $x_p$ values from the simulated \lya\ profiles.}  
\label{fig:xp_NHI}
\end{figure}

\section{Method}
\label{sec:Method}
\subsection{Identification of halos}
We identify halos in the simulation snapshots using the halo finder code {\sc Rockstar-Galaxies} \citep{rockstar2013}. A structure is identified as a halo if it contains at least 30 dark matter particles and 10 star particles within its virial radius. As shown in Table~\ref{tab:M_star_list}, the number ($N_h$) and median stellar mass ($M_*$) of the identified halos depend on the simulation volume and mass resolution.

As a validation, we compare the galaxy stellar mass function (GSMF) from  our  simulation boxes at two different redshifts with the observation in Figure~\ref{fig:GSMF}. Halo stellar masses are computed by summing the masses of all star particles within the virial radius. The GSMF obtained here is consistent with that obtained by combining the halos in three {\sc Ninja} simulation boxes with side-lengths of 50, 150 and 250~$h^{-1}$cMpc and $2\times2040^3$ DM+gas particles \citep[for details, see ][]{Behera2026}. The GSMF observations are taken from both pre-JWST era observations such as the HST imaging and the Spitzer/IRAC data \citep{song2016, stefanon2021} and from the JWST COSMOS-Web survey in combination with multiband photometric ancillary imaging \citep{harvey2025, shuntov2025}. Although our simulations under-predict the observed GSMFs, \citet{Behera2026} showed that they reproduce the observed UV luminosity functions well. The discrepancy likely arises because observational GSMFs are derived from SED fitting, whereas the simulations use the total stellar mass within each halo. A detailed analysis will be presented in a future {\sc Ninja} paper. The GSMF from 50h$^{-1}$cMpc boxes for different resolutions are consistent with each other and with the results from high resolution runs in the overlapping mass range. However, the GSMF for 140h$^{-1}$cMpc box is slightly lower compared to the other boxes, indicating possible convergence issues.

\subsection{Simulated absorption spectra}
\label{subsec:sim_abs_spec}

Our aim is to simulate the \lya\ absorption profile of galaxies and compare them with the recent JWST observations. We generate three sets of spectra  considering the contribution from the IGM, galaxy+IGM and residual hydrogen in ISM within the \HII\ region+galaxy+IGM to the \lya\ profile, as follows.

For Case-I, we consider density, temperature, and velocity fields along the random direction to generate an absorption profile using the standard procedure used for the \lya\ forest studies \citep[for example as explained in][]{Mallik2023}. We keep the \xhi\ as a free parameter. We use these spectra to quantify the expected Gunn-Peterson absorption  from the IGM with an uniform \xhi. 
Sightlines of length \(225\,h^{-1}\) cMpc are constructed by stitching together random segments through the simulation box.

For the Case-II, we identify the halos (as described in the previous section) for the simulation snapshots at $z$ = 8, and 10. For each halo, the sightlines are shot towards the center of mass of the star particles associated with that halo, along $\pm x$, $\pm y$, and $\pm z$ axes, i.e., we shoot six sightlines per halo. These sightlines start from the halo center and continue for half the side-length of the simulation box, i.e., 70$h^{-1}$ cMpc for the "L140N1008" box and 25$h^{-1}$cMpc for the L50N1008 and L50N640 boxes. To account for the contribution of neutral gas at a large distance from the halos, we stitch IGM sightlines at the end of these sightlines and truncate the sightlines at 225$h^{-1}$~cMpc for all three boxes.

Since our simulations do not model the effects of reionization processes, we consider a simple toy-model in postprocessing to account for the ionization of the gas surrounding the halos. As shown in the left panel of Figure~\ref{fig:model1}, around each galaxy, we consider an ionized spherical bubble of radius $R_{\rm HII}$ inside which the \HI\ gas is assumed to be completely ionized. While generating the spectra, we vary $R_{\rm HII}$ within the range $0~\mathrm{pkpc} \le R_{\hii} \le 400~\mathrm{pkpc}$. This $R_{\mathrm{HII}}$ range is consistent with the ionized bubble radius from theoretical exceptions described in Appendix~\ref{sec:galactic_emission}. The neutral hydrogen fraction ($x_{\mathrm{HI}}$) in the medium outside (i.e general IGM) the bubble is $0.1$, $0.8$ or $1.0$. Note that varying $R_{\rm HII}$ is like varying the escape fraction of the Lyman continuum (LyC) photons and/or intrinsic reddening of the photons from the galaxy for a given escape fraction. Varying \xhi\ outside the ionized bubble is done to mimic the galaxy being formed in a region that already has partial ionization. In this model (when  $R_{\rm HII} \ne 0$) we ignore any intrinsic damped \lya\ absorption originating from the ISM or the circumgalactic medium (CGM) of the galaxies.

In Case III, we investigate the effect of residual neutral hydrogen within the ionized bubble. For this purpose, we adopt a modified version of Case II, illustrated in the right panel of Figure~\ref{fig:model1}. In this model, the high-density gas within the virial radius ($R_{\mathrm{vir}}$) of the halos is assumed to have $x_{\rm{HI}}^V$,
while the gas between $R_{\mathrm{vir}}$ and the ionized bubble radius $(R_{\mathrm{HII}})$ is assumed to be fully ionized. We discuss the results for two scenarios with $x_{\rm{HI}}^V$=~0.3  or 1 and $x_{\rm{HI}}$=0.8 or 1.0 in section~\ref{subsec:caseIII}. Similar to the sightlines considered in Case I and Case II, the length of the sightlines considered in Case III is $225h^{-1}$~cMpc. 

For simplicity, we assume the galaxy spectrum to be flat without including the \lya\ emission line. We generate the density ($n_H$), temperature ($T$), and peculiar velocity ($v$) fields at the grid points along a sightline towards a halo using smoothed-particle-hydrodynamics (SPH) smoothing of these fields of the gas particles within smoothing length. In the upper three panels of Figure \ref{fig:demo_los}, we show these fields for an example sightline around a halo with halo mass of $10^{11.4}M_\odot$ and stellar mass of $10^{9.25}M_\odot$ for case II configuration. We generate an optical depth profile following standard procedure that takes into account natural and thermal broadening and velocity shifts due to cosmic expansion and peculiar velocities. We compare the spectra generated for different values of $R_{\rm HII}$ and $x_{\mathrm{HI}}$ in panels (d) and (e), respectively.
To account for the low spectral resolution of the NIRSpec PRISM mode (R $\sim 100$ for wavelength $0.6-5.3 ~\mu m$), we convolve the simulated spectra with a Gaussian profile with FWHM of 3000~\kms. We also add the Poissonian noise corresponding to an SNR of 30 to each pixel of the mock spectra. Panels (d) and (e) in Figure~~\ref{fig:demo_los} show the \lya\ profiles before and after applying instrumental broadening in faint and bright colors respectively. 

\subsection{Quantifying the strength of the absorption}
\label{subsec:xp_quantify}

We quantify the strength of the damped \lya\ absorption through a parameter, $x_p$, defined as the velocity with respect to the galaxies (i.e. halos) at which the normalized transmitted flux reaches 10$\%$, assuming the redward of the source to have a positive value of $x_p$. 
The horizontal dashed line in panels (d) and (e) shows the transmitted flux of 0.1. In panel (d) of  Figure~\ref{fig:demo_los} we show the \lya\ absorption profile around a chosen halo for $R_{\hii}~=~0~\mathrm{pkpc}$ and $x_{\mathrm{HI}}$ = 1 and 0.1. As panel (d) mentions, the $x_p$ value for $x_{\mathrm{HI}}=1.0$ spectra is significantly higher compared to the spectra for $x_{\mathrm{HI}}$ = 0.1.  In panel (e) we show the effect of $R_\hii$, using absorption spectra obtained assuming $R_{\hii}=0~\mathrm{pkpc}$ and $R_{\hii}=400~\mathrm{pkpc}$. The absorption profile for $R_{\hii}=400~\mathrm{pkpc}$ is significantly shallower than the profile for $R_{\hii}=0~\mathrm{pkpc}$ for the same $x_{\mathrm{HI}}$ values. This figure clearly demonstrates that the observed \lya\ profile will be sensitive to both the size of the H~{\sc ii} region and the the overall neutral fraction of the IGM.

We next associate an equivalent \HI\ column density of the gas producing the \lya\ absorption with $x_p$ for enabling comparison with observational results. We generate absorption profiles arising from a single cloud with \HI\ column density within the range $10^{18}-10^{23} ~\mathrm{cm}^{-2}$, and calculate the $x_p$ values for these profiles. The column density, $N$(\HI), and corresponding $x_p$ values of these profiles are plotted in Figure~\ref{fig:xp_NHI}, both before (in blue-colored lines) and after (in orange-colored lines) considering the effects of the instrumental resolution, for Doppler parameter values (b) of 50~\kms, 100~\kms, 150~\kms, 200~\kms. Due to the low resolution of the JWST NIRSpec PRISM mode observation, the $x_p$ values of absorption profiles corresponding to \HI\ column density less than $2.57 \times 10^{21} ~\mathrm{cm}^{-2}$ are negative, as indicated by the red dotted line in Figure~\ref{fig:xp_NHI}. We quantify the equivalent \HI\ column density of the absorbing gas from the $x_p$ values of the simulated spectra by interpolating the tabulated $x_p$ - $N$(\HI) values.  As indicated in Figure~\ref{fig:xp_NHI}, the value of the $b$ parameter does not change the $x_p$ values significantly for a given column density for $N$(\HI) $>10^{21}~\mathrm{cm}^{-2}$. Hence we used $b = 100$~\kms\ to estimate the equivalent \HI\ column densities.

\section{Results}
\label{Sec:results}

In this section, we compare the properties of IGM and halo sightlines across different simulation boxes and examine the resulting Ly$\alpha$ absorption profiles. We also assess our predictions against recent observational constraints.

Our primary comparison is with \citet{Heintz2025}, who analyzed 494 galaxies from the JWST-PRIMAL survey at $z=5.5$--13.4. They inferred \NHI\ from rest-frame equivalent widths measured over 1180--1350~\AA\ and found that $\sim65$--90\% of galaxies at $z>8$ have $N_{\mathrm{HI}}>10^{21}~\mathrm{cm}^{-2}$. This serves as one of the key observational constraints for our analysis.
We also compare our results with \citet{Mason2025}, who modeled 99 JWST/NIRSpec PRISM galaxies at $z=5.5$--13 using the simulations of \citet{Lu2024} and forward spectral modeling. Their framework includes H~{\sc i} absorption from the galaxy, the surrounding H~{\sc ii} region, and the neutral IGM, with the IGM contribution depending on the H~{\sc ii} region size and the IGM neutral fraction. Marginalizing over model parameters, they inferred a median intrinsic absorber column density of $\log(N_{\mathrm{HI}}/\mathrm{cm}^{-2})=20.8$, found that only $\sim20\%$ of sightlines show absorption stronger than expected from the IGM alone, and estimated that $\sim18\%$ of galaxies host intrinsic damped Ly$\alpha$ absorbers with $\log(N_{\mathrm{HI}}/\mathrm{cm}^{-2})\ge22$ (i.e $f_{22} \sim 18\%)$. Since our analysis measures the total H~{\sc i} column density, we treat these intrinsic values as lower limits when comparing with our simulations.

Finally, we consider the results of \citet{Pollock2026}, who reported median upper limits on \NHI\ of $10^{21.71}$, $10^{22.24}$, and $10^{22.34}~\mathrm{cm}^{-2}$  over the redshift ranges $z=9$--10, 10--12, and $>12$, respectively. They also inferred IGM-dominated fractions of 0.42, 0.50, and 0.29 across these intervals using a two-component (IGM + intrinsic DLA) fit to the Ly$\alpha$ profile.
We evaluate these observational inferences in the context of our simulation results.

\subsection{Radial density profile around the halos and IGM}
\label{sec:rad_profile}
Before getting into to the DLA profiles, we investigate the gas over-density as a function of radial distance from the identified halo centre of mass. For this analysis, we use the SPH-smoothed over-density values computed on the grids along the sightlines. As the halo mass distributions in the $50~h^{-1}\mathrm{cMpc}$ and $140~h^{-1}\mathrm{cMpc}$ simulation boxes differ significantly, we first restrict the comparison to a common halo mass range. At $z\sim 8$, most halos in the $50~h^{-1}\mathrm{cMpc}$ boxes have halo-masses in the range $10^{9.5}$--$10^{10.5}~M_\odot$, whereas 98\% of the halos in the $140~h^{-1}\mathrm{cMpc}$ box lie in the halo-mass range $10^{10.5}$--$10^{11.5}~M_\odot$. Specifically, only 148 (143) of the 772 (555) halos in the L50N1008 (L50N640) simulation fall within the mass range $10^{10.5}$--$10^{11.5}~M_\odot$, while the L140N1008 simulation contains 1197 halos in the same mass range. As all three simulations contain more than 100 halos in this common mass range, we use these halos to compare the radial over-density profiles and assess their numerical convergence. 

\begin{figure}[htbp]
    \begin{minipage}{0.48\columnwidth}
        \centering
        \includegraphics[width=2\linewidth]{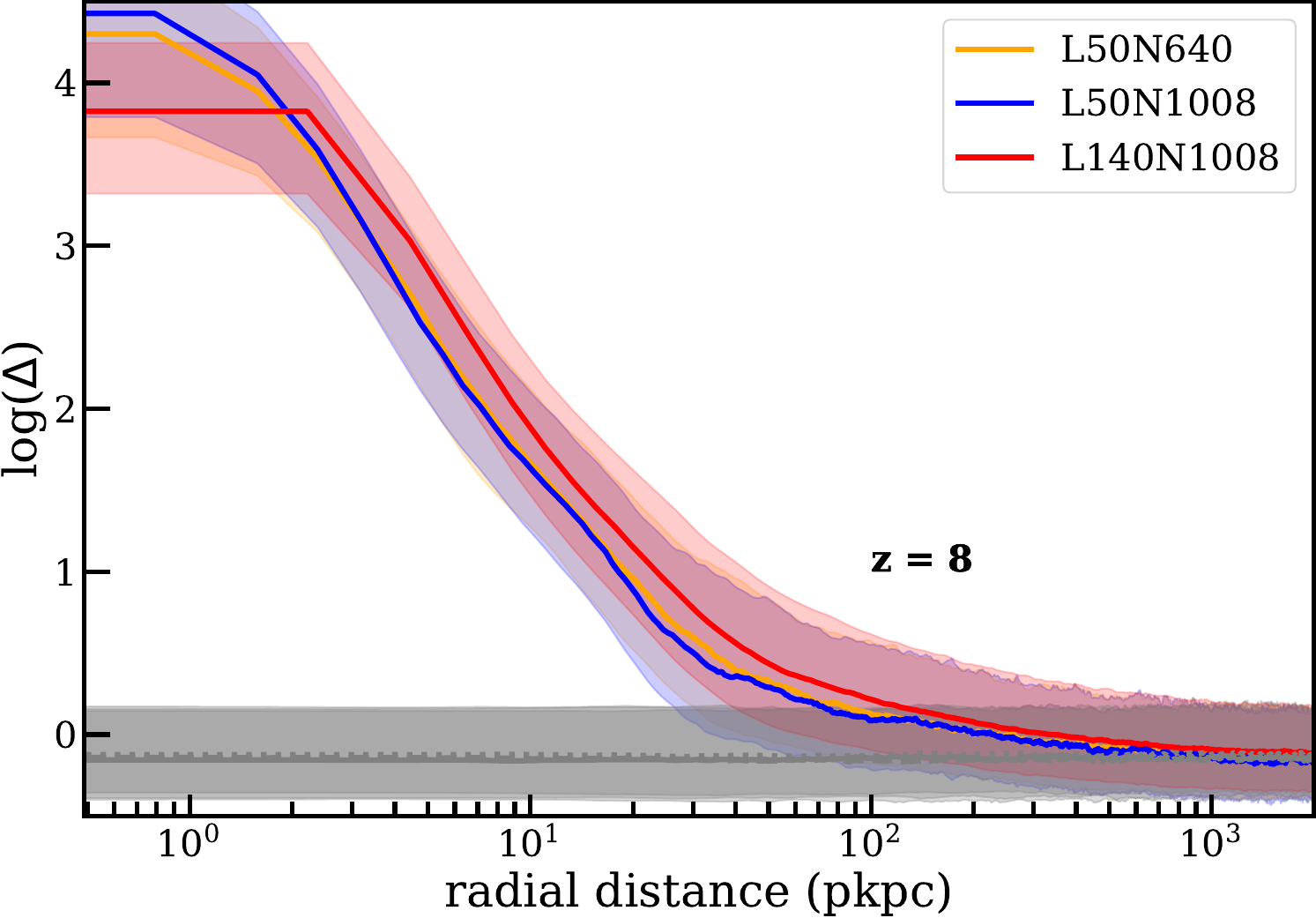}
    \end{minipage}
    \caption{Comparison of radial over-density profile for halos within halo mass range $10^{10.5} - 10^{11.5}M_\odot$ at $z=8$.
    The orange, blue and red colored lines are for boxes  L50N640, L50N1008 and L140N1008, respectively. The IGM over-density for boxes  L50N640, L50N1008 and L140N1008 are shown in solid, dashed and dotted gray lines. }
    \label{fig:rad_profile2}
\end{figure}

The radial gas over-density profiles around halos (at $z=8$) in this mass range are shown in Figure~\ref{fig:rad_profile2} for the three simulations. For comparison, the median over-density measured along 1000 randomly selected IGM sightlines is shown by the gray curve. In both cases, the shaded regions indicate the $1\sigma$ scatter. As expected, the over-density distribution along the IGM sightlines is well converged.
The gas over-density profiles around halos of similar masses are in good agreement among the three simulations, with differences remaining within the quoted scatter. This demonstrates a good numerical convergence of the radial profiles. In all three simulations, the gas over-density around halos remains significantly higher than that along the IGM sightlines out to radial distances of $\sim100$ pkpc.
In Figure~\ref{fig:rad_profile1}, we show the median gas over-density as a function of radial distance for all halos with stellar masses greater than $10^7~M_\odot$ in the three simulations and for  $z=10$ and $z=8$.

For the L50N1008 simulation, the median over-density along the halo sightlines decreases to the upper $1\sigma$ limit of the IGM over-density distribution at a radial distance of $R_{eq}\sim66$~pkpc at both $z=10$ and $z=8$, indicating little or no evolution in this characteristic scale between the two redshifts. A similar behaviour is found in the L50N640 and L140N1008 simulations, where the transition radius ($R_{eq}$), beyond which the halo over-density becomes comparable to that of the IGM, also shows no significant evolution with redshift. At a fixed redshift, however, the transition radius is larger in the L140N1008 simulation ($R_{eq}\sim 105$~pkpc) than in the L50N1008 simulation. This difference is likely driven by the dependence of the radial gas over-density profile on halo mass, since the characteristic halo masses differ between the simulation boxes (see Section~\ref{sec:mass-HI-relation} ). 
\begin{figure*}
\begin{minipage}{\textwidth}
\centering
\includegraphics[width=\textwidth]{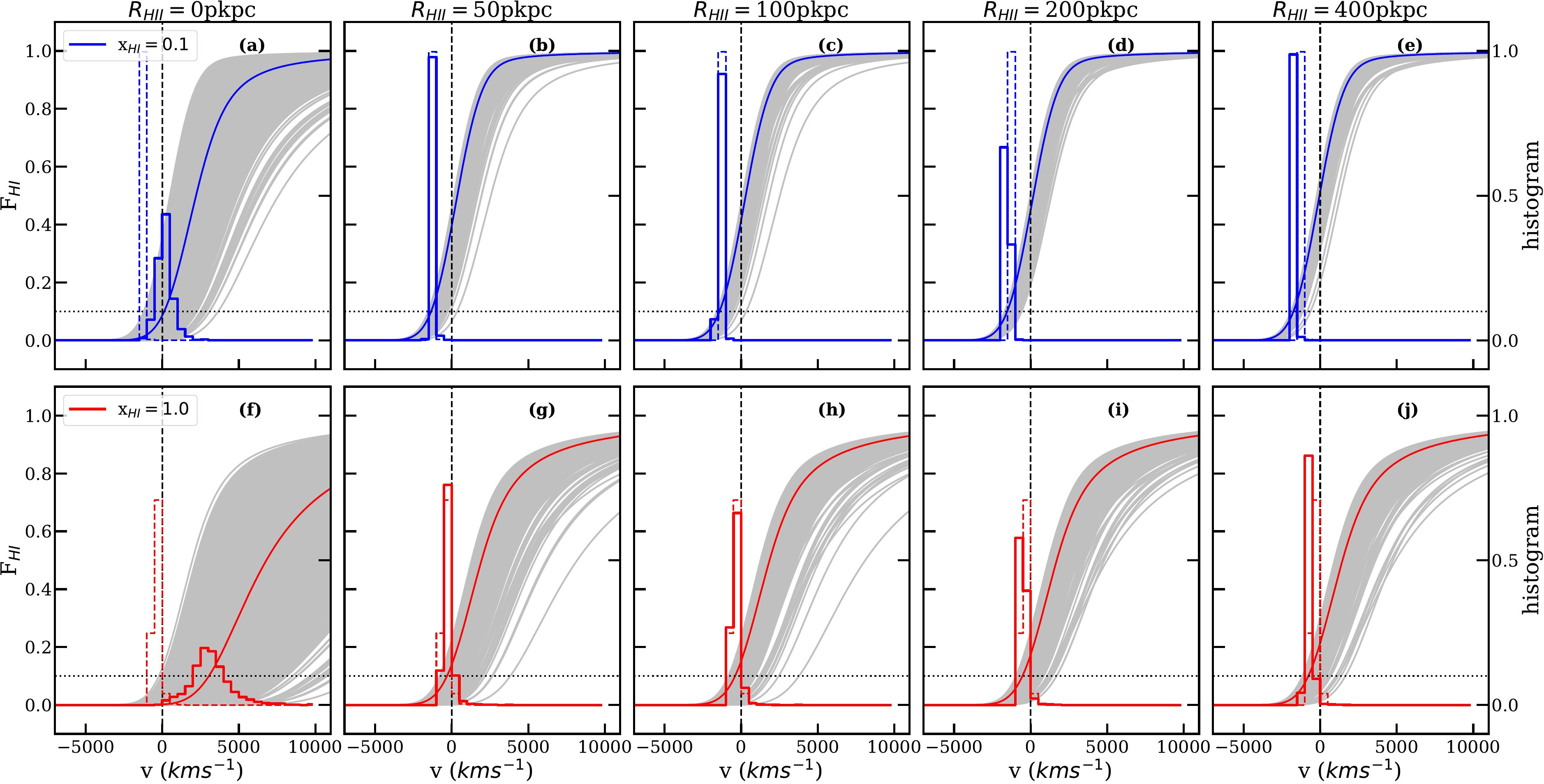}
	\caption{The noise free \lya\ absorption profiles (shown in gray for case~II) towards the halo at $z=8$s for varied $R_{\rm HII}$ and \xhi in L50N1008. The upper and lower panels show the results for the \xhi\ = 0.1 and 1.0, respectively. The median absorption profiles for each panel are shown in red. The histogram for $x_p$ distribution corresponding to each scenario is also shown for halo(IGM) sightlines in solid(dashed) lines. It is evident that Case-II produces stronger absorption for  $R_{\rm HII}<100$ pkpc compared to Case-I.
	}
\label{fig:profiles}
\end{minipage}%
\end{figure*}

\subsection{The \lya\ profile $x_p$ values in the IGM and halo sightlines}
\label{subsec:DLA_result}

In this section, we focus on the \lya\ absorption profile and the effective \NHI\ from halos in L50N1008 box.
For this, we construct the Ly$\alpha$ absorption profiles along randomly selected IGM sightlines and sightlines towards the centre of mass of the halos (i.e galaxy or halo sightlines) following the procedure described in Section~\ref{subsec:sim_abs_spec}. Figure~\ref{fig:profiles} shows the Ly$\alpha$ profiles for sightlines towards the identified halos at $z=8$, plotted as gray curves. The upper and lower panels are for $x_{\mathrm{HI}}$ = 0.1 and 1.0, respectively. From left to right, the panels show models with increasing ionized bubble radii covering $0\le R_{\rm HII}\le 400$ pkpc.
%
As expected, for a given value of $R_{\rm HII}$, the Ly$\alpha$ absorption is significantly stronger for $x_{\mathrm{HI}}$ = 1.0 than for $x_{\mathrm{HI}}$ = 0.1
The absorption profiles become progressively shallower with increasing $R_{\rm HII}$, reflecting the reduced contribution of nearby neutral gas to the Ly$\alpha$ damping wing.

We determine the $x_p$ values for each sightline following the procedure described in Section~\ref{subsec:xp_quantify}. The resulting $x_p$ distributions for the halo sightlines are shown as solid histograms in each panel, while the corresponding distributions for randomly selected IGM sightlines are shown by the dashed histograms.
We find that the median $x_p$ values for sightlines towards halos are systematically higher than those for the IGM sightlines when $R_{\rm HII} < 100$ pkpc. In addition, the median $x_p$ value for $R_{\rm HII}=0$ pkpc is significantly larger than those obtained for larger $R_{\rm HII}$, highlighting the strong influence of H~{\sc i} in the immediate vicinity of halos. The broad $x_p$ distribution for $R_{\rm HII}=0$ pkpc further suggests that the Ly$\alpha$ absorption profiles around halos vary substantially with direction. The spread in $x_p$ also indicate a dependence of $x_p$ on the halo mass and/or stellar mass of the halos.
 
We characterize the Ly$\alpha$ absorption in these sightlines using three parameters: the percentage of absorbers with inferred \NHI\ greater than $10^{21}~\mathrm{cm}^{-2}$ (denoted by $f_{21}$), the median \NHI\ of these absorbers (denoted by $\langle N_{\hi} \rangle)$, and the percentage of absorbers exhibiting a \lya\ damping wing signature, i.e., those with column densities exceeding $10^{20.3}~\mathrm{cm}^{-2}$ (denoted by $f_{\rm DLA}$).

\subsection{Contribution of IGM to the damped-\lya\ profile (Case~I)}

\begingroup
\setlength{\tabcolsep}{4pt} 
\begin{table}
    \centering
    \small
    \caption{Different parameters characterizing the absorbers (see Section~\ref{subsec:DLA_result} for their definition) in the IGM sightlines (case~I) in the three simulation boxes.
    }
     \begin{threeparttable}
\begin{tabular}{
>{\centering\arraybackslash}p{0.5cm}|
>{\centering\arraybackslash}p{0.20cm}
>{\centering\arraybackslash}p{0.5cm}
>{\centering\arraybackslash}p{1.2cm}|
>{\centering\arraybackslash}p{0.20cm}
>{\centering\arraybackslash}p{0.5cm}
>{\centering\arraybackslash}p{1.2cm}|
>{\centering\arraybackslash}p{0.20cm}
>{\centering\arraybackslash}p{0.5cm}
>{\centering\arraybackslash}p{1.2cm}
}
   \hline
   & \multicolumn{3}{c|}{L50N1008} & \multicolumn{3}{c|}{L50N640} & \multicolumn{3}{c}{L140N1008} \\

     \xhi & $f_{21}$ & $f_{\rm DLA}$ & $\log\langle N_{\hi}\rangle$  & $f_{21}$ & $f_{\rm DLA}$ & 
     $\log\langle N_{\hi}\rangle$  & $f_{21}$ & $f_{\rm DLA}$ & $\log\langle N_{\hi}\rangle$  \\
     \hline
\multicolumn{10}{c}{$z=8$} \\
0.1 & <1 & <1 & - & 0 & <1 & - & 0 & <1 & - \\

0.8 & 38 & 100 & 20.95 & 39 & 100 & 20.96 & 44 & 100 & 20.98 \\

1.0 & 81 & 100 & 21.11 & 84 & 100 & 21.12 & 87 & 100 & 21.13\\
\multicolumn{10}{c}{$z=10$} \\
0.1 & 0 & <1 & - & 0 & <1 & - & 0 & <1 & - \\

0.8 & 97 & 100 & 21.19 & 97 & 100 & 21.19 & 98 & 100 & 21.21\\

1.0 & 100 & 100 & 21.34 & 100 & 100 & 21.34 & 100 & 100 & 21.36\\
\hline
    \end{tabular}
    \label{table:xp_z_IGM}
     \end{threeparttable}
\end{table}
\endgroup
In Table~\ref{table:xp_z_IGM}, we report $f_{21}$, $f_{\rm DLA}$, and $\log \langle N_{\hi} \rangle$ for case~I, where only H~{\sc i} absorption from the IGM is included, for all three simulation boxes at $z=8$ (upper panel) and $z=10$ (lower panel). For a given combination of $z$ and $x_{\mathrm{HI}}$, all three quantities agree to within $\sim1\%$ between the L50N1008 and L50N640 simulations. In contrast, the L140N1008 box yields systematically higher $f_{21}$ values, by up to $\sim6\%$, than the L50N1008 simulation. 
At both $z=8$ and $z=10$, no sightlines produce Ly$\alpha$ damping wings corresponding to $N$(H~{\sc i}) $> 2 \times 10^{20}~\mathrm{cm}^{-2}$ for $x_{\mathrm{HI}}=0.1$. This contrasts with the frequent detection of strong Ly$\alpha$ damping wings in $z>8$ JWST galaxies, implying that the average neutral fraction of the intervening IGM is substantially higher than 0.1, consistent with the reionization models discussed in the Introduction. In contrast, all spectra exhibit DLA absorption for $x_{\mathrm{HI}}\ge0.8$.

The fraction $f_{21}$ depends strongly on both $x_{\mathrm{HI}}$ and redshift. At $z\sim8$, for $x_{\mathrm{HI}}=0.8$, 38\% and 44\% of the spectra from the L50N1008 and L140N1008 boxes, respectively, are consistent with $N_{\hi}>10^{21}~\mathrm{cm}^{-2}$. These fractions increase to 81\% and 87\% for $x_{\mathrm{HI}}=1.0$. At $z\sim10$, the $f_{21}$ values are nearly identical for the L50N1008 and L50N640 simulations: $\sim97\%$ of the spectra satisfy $N_{\hi}>10^{21}~\mathrm{cm}^{-2}$ for $x_{\mathrm{HI}}=0.8$, increasing to 100\% for $x_{\mathrm{HI}}=1.0$. For $z\sim8$, we do see $f_{21}$ in L140N1008 box is slightly higher but consistent within 6\% of that found for L50N1008. However, the convergence is $\le1\%$ for $z\sim 10$.

Since the predicted values $f_{21}$ for $x_{\mathrm{HI}}\le0.8$ are lower than the fraction reported by \citet{Heintz2024}, observations would favor $x_{\mathrm{HI}}>0.8$ if the damping wings arise solely from neutral hydrogen in the IGM. However, detection of a small fraction of galaxies at $z\sim$10 without a strong \lya\ damping wing will require large H~{\sc ii} regions around galaxies.
On the other hand, even for $x_{\mathrm{HI}}=1$, none of the simulated IGM sightlines reaches $N_{\hi}>10^{22}~\mathrm{cm}^{-2}$. As noted by \citet{Mason2025}, IGM absorption alone therefore cannot reproduce the substantial fraction of JWST galaxies with Ly$\alpha$ absorption profiles consistent with $N_{\hi}>10^{22}~\mathrm{cm}^{-2}$ \citep[e.g.,][]{Heintz2025}. Explaining these systems requires an additional contribution from gas within and around galaxies (i.e., halos). This is what we explore next.

\subsection{Contribution of galaxy + IGM to the \lya\ profile ({\rm case-II})}
\begingroup
\setlength{\tabcolsep}{4pt} 
\begin{table*}
    \centering
    \small
    \caption{Different parameters characterizing the absorbers in the galaxy sightlines (case~II) at z=8 for the two $50 h^{-1}$cMpc boxes. The description of $f_{21}$, $f_{\rm DLA}$ and log~$\langle N_{\hi}\rangle$ are same as Table~\ref{table:xp_z_IGM}. }
     \begin{threeparttable}
   \begin{tabular}{c|ccc|ccc|ccc|ccc|ccc}
   \hline
   & \multicolumn{3}{c|}{$R_{\rm HII}=0$ pkpc} & \multicolumn{3}{c|}{$R_{\rm HII}=50$pkpc} & \multicolumn{3}{c|}{$R_{\rm HII}=100$ pkpc} & \multicolumn{3}{c|}{$R_{\rm HII}=200$pkpc} & \multicolumn{3}{c}{$R_{\rm HII}=400$ pkpc} \\

     \xhi & $f_{21}$  & $f_{\rm DLA}$  & log~$\langle N_{\hi}\rangle $  & $f_{21}$ & $f_{\rm DLA}$ & log~$\langle N_{\hi}\rangle $  & $f_{21}$ & $f_{\rm DLA}$ & log~$\langle N_{\hi}\rangle $  & $f_{21}$ & $f_{\rm DLA}$ & log~$\langle N_{\hi}\rangle $  & $f_{21}$ & $f_{\rm DLA}$ & log~$\langle N_{\hi}\rangle $   \\
     \hline

\multicolumn{16}{c}{For L50N1008 at z=8} \\
0.1 & 92 & 99 & 21.49 & <1 & 2 & 19.11 & <1 & 1 & 19.08 & 0 & <1 & - & 0 & <1 & -\\

0.8 & 100 & 100 & 22.42 & 60 & 100 & 21.04 & 41 & 100 & 20.96 & 17 & 99 & 20.81 & 3 & 77 & 20.49\\

1.0 & 100 & 100 & 22.52 & 91 & 100 & 21.19 & 78 & 100 & 21.11 & 47 & 100 & 20.98 & 11 & 98 & 20.73\\
\multicolumn{16}{c}{For L50N640 at z=8} \\
0.1 & 92 & 100 & 21.41 & <1 & 3 & 19.09 & <1 & 1 & 19.06 & 0 & <1 & - & 0 & 0 & -\\
0.8 & 100 & 100 & 22.34 & 63 & 100 & 21.04 & 42 & 100 & 20.96 & 17 & 100 & 20.81 & 3 & 79 & 20.50\\
1.0 & 100 & 100 & 22.44 & 93 & 100 & 21.19 & 79 & 100 & 21.12 & 49 & 100 & 20.99 & 11 & 98 & 20.74\\
\hline

    \end{tabular}
    \label{table:xp_z8_gal}
     \end{threeparttable}
\end{table*}
\endgroup
{

Here, we assume each halo to be surrounded by a fully ionized (i.e $x_{\mathrm{HI}} = 0$) spherical H~{\sc ii} region  of radius $R_{\hii}$. Therefore, depending upon the value of $R_{\hii}$, the damped Ly$\alpha$ absorption along these sightlines receives contributions from the ISM, the over-dense gas surrounding halos (discussed in Section~\ref{sec:rad_profile}), and the IGM.
For simplicity, we assume that the gas in these three components follows the same prescribed neutral fraction, $x_{\mathrm{HI}}$, when they are outside $R_{\hii}$ (see the left panel of Figure~\ref{fig:model1}). The resulting damped Ly$\alpha$ profile generation is therefore characterized by two parameters: $x_{\mathrm{HI}}$ and $R_{\hii}$.}

In Tables~\ref{table:xp_z8_gal} and \ref{table:xp_z10_gal}, we report the parameters characterizing the absorbers for Case~II in the two 50$h^{-1}$cMpc boxes at $z=8$ and $z=10$, respectively. The three quantities reported for a given combination of $x_{\mathrm{HI}}$ and $R_{\hii}$ show a good convergence between the two simulations.
For all three values of $x_{\mathrm{HI}}$, the $f_{21}$ and $\langle N_{\hi}\rangle$ values at $R_{\hii}=100$ pkpc becomes comparable to those of the IGM sightlines with similar constant $x_{\mathrm{HI}}$. 
For larger values of $R_{\hii}$, the absorption becomes weaker than that predicted in Case~I. This behaviour is expected, as the characteristic radii $R_{\rm eq}$ in our models typically lie between 50 and 100 pkpc (see Figure~\ref{fig:rad_profile2}). We find that the values of $f_{21}$ and $f_{\rm DLA}$ for Ly$\alpha$ absorbers decrease systematically with increasing $R_{\hii}$ for all values of $x_{\mathrm{HI}}$.

The limiting case of $R_{\hii}=0$ pkpc and $x_{\mathrm{HI}}=1$ represents a galaxy embedded in a fully neutral IGM, with the line of sight probing fully neutral gas in the ISM, the surrounding overdense regions, and the IGM. Our models predict that more than 90\% of galaxies will exhibit $\log N(\HI) > 21$ (i.e., $f_{21}>90\%$) for $x_{\mathrm{HI}}\ge0.1$. This predicted value exceeds the range reported by \citet{Heintz2025}, suggesting that at least a substantial fraction of galaxies must be surrounded by ionized bubbles, i.e., $R_{\hii}$ must be non-zero for a significant fraction of the galaxy population.
The predicted values of $f_{22}$ further support this conclusion. Our models yield $f_{22}$ values of 17\%, 94\%, and 96\% for $x_{\mathrm{HI}}=0.1$, 0.8, and 1.0, respectively. While the prediction for $x_{\mathrm{HI}}=0.1$ is consistent with the values inferred by \citet{Mason2025}, the predictions for $x_{\mathrm{HI}}=0.8$ and 1.0 are substantially higher than the inferred values.
Taken together with the results presented in Table~\ref{table:xp_z_IGM}, these comparisons indicate that, if the IGM at $z\sim8$ is highly neutral ($x_{\mathrm{HI}}\gtrsim0.8$), the gas in the ISM and the immediate surroundings of galaxies must be substantially ionized. In particular, the observations require local ionized regions around at least a significant fraction of galaxies to suppress the incidence of high H~{\sc i} column densities.  From Table~\ref{table:xp_z10_gal}, we can infer that the same conclusions are valid for $z\sim 10$ as well.

Next, we consider a gas with  $x_{\mathrm{HI}}=0.8$ and 1.0 surrounding the H~{\sc ii} regions with radius $\ge50$ pkpc. For $z\sim8$,  $R_{\hii}<200$ pkpc for $x_{\mathrm{HI}}=1.0$ (and 100 pkpc for $x_{\mathrm{HI}}=0.8$) is able to produce $f_{21}$ in the range of 65-90\% observed by \citet{Heintz2025} (see Table~\ref{table:xp_z8_gal}). 
For $z\sim10$,  $R_{\rm HII}<400$ pkpc for ¸ (and $<$200 pkpc for $x_{\mathrm{HI}}=0.8$) is able to produce $f_{21}$ in the range (i.e 65-90\%) observed by \citet{Heintz2025}.
However, when $R_{\hii} \ge50$ pkpc we find $f_{22}$ to be less than 1\% even for  $x_{\mathrm{HI}}=1.0$. This once again confirms the earlier finding that, to produce strong damped \lya\ systems with $\log N(\HI) \ge 22$ our models require contributions of neutral gas from the ISM {and over-dense regions around galaxies}.

Comparing the values $f_{21}$ and $f_{\rm DLA}$ parameters between $z=10$ and $8$, we find that there is a trend of stronger absorption at higher redshift (see Tabels~\ref{table:xp_z8_gal} and \ref{table:xp_z10_gal}). The value of $\langle N_{\hi}\rangle $ between the two redshifts differs by $ \sim 0.2$ dex for $R_{\hii}\leq 100$ pkpc. For $z\sim10$, the values of $f_{21}$ for $R_{\hii} \le 50~\mathrm{pkpc}$ for \xhi$\ge 0.8$ is higher than the upper bound of the observed $f_{21}$ range reported in \citet{Heintz2025}. It is also evident that for \xhi\ = 1 (respectively \xhi\ = 0.8) we need $200\le R_{\hii} ({\rm pkpc})\le 400$ (respectively  $100\le R_{\hii} ({\rm pkpc})\le 200$)
to produce the range of $f_{21}$ consistent with the observations of \citet{Heintz2025}. For $z\sim10$ also, we find that to produce the observed value of $f_{22}$ we need an additional contribution to the neutral hydrogen column density coming from the ISM gas.
Similar to the case-I, the values of $f_{21}$, $f_{\rm DLA}$ and $\langle N_{\hi}\rangle $ of the \lya\ absorbers in the L50N640 box for case-II are also very close to the values obtained for the L50N1008 box. Due to the difference in the range of stellar masses of the halos and the particle resolution between the L50N1008 and L140N1008 boxes, we will compare the values of $f_{21}$, $f_{\rm DLA}$ and $\langle N_{\hi}\rangle $ of the \lya\ absorbers between these two boxes using a subset of sightlines shoot around the halos within similar mass range in section \ref{subsec:comparison}. 

In summary, in Case II models without H~{\sc ii} regions around galaxies predict much stronger \lya\ absorption than observed, indicating that H~{\sc ii} regions are present around at least a significant fraction of halos. Conversely, models with fully ionized H~{\sc ii} regions larger than $\sim$50 pkpc fail to reproduce the observed strong damped \lya\ absorption, suggesting that an additional contribution from residual neutral H~{\sc i} within the H~{\sc ii} regions is required when their sizes approach or exceed $R_{eq}$.

\subsection{Contribution of residual hydrogen in ISM within the \hii\ region+galaxy+IGM to the \lya\ profile (case-III)}
\label{subsec:caseIII}
{
To address the shortcomings of Case II, we consider a more realistic configuration in Case III. Here, each sightline originates at the halo centre, traverses partially ionized gas within  the virial radius ($R_{\rm vir}$) of the halo (having a neutral fraction of $x_{\mathrm{HI}}^{\rm V}$), passes through fully ionized gas between $R_{\rm vir}$ and $R_{\hii}$, and finally intersects the surrounding IGM (Figure~\ref{fig:model1}), where the neutral fraction is $x_{\mathrm{HI}}$. The predicted \lya\ profile therefore depends on three free parameters: $x_{\mathrm{HI}}^{\rm V}$, \xhi, and $R_{\rm HII}$. For simplicity, we present the results primarily for the L50N1008 simulation at $z\sim8$.

We first consider the  case with $x_{\mathrm{HI}}^{\rm V}$ = \xhi\ = 1 and $R_{\hii}=50$ pkpc.  In this case, we obtain $f_{21}=91\%$, comparable to the value found for Case II (Table~\ref{table:xp_z8_gal}), while $f_{22}$ increases only marginally to 3\% (from $\leq1\%$). Thus, for halos in the L50N1008 simulation, neutral gas confined within the virial radius alone is insufficient to produce strong damped \lya\ absorption when the surrounding over-dense regions are substantially ionized. In contrast, in Case II we obtained $f_{22}=17\%$ even for \xhi$=0.1$ at $R_{\hii}=0$ pkpc, which is significantly higher than the value found here despite assuming completely neutral gas within $R_{\rm vir}$. This suggests the important role of residual neutral gas in the over-dense regions surrounding halos, at least over the halo mass range probed by our simulations, in producing strong \lya\ damping-wing absorption.

An alternative explanation for the low values of \(f_{22}\) in Case III is that the halos in our simulations are systematically less massive than those hosting the observed galaxies. Since high-resolution RT simulations predict that \(N_{\hi}\) correlates with both stellar and halo mass \citep[e.g.,][]{Gelli2025,Steen2026}, we repeated the analysis for halos with masses in the range $10^{10.5}-10^{11.5}{\rm M_\odot}$. For \(x_{\mathrm{HI}}^{\rm V}=x_{\hi}=1\) and \(R_{\rm HII}=50\) pkpc, we obtain \(f_{21}=94\%\) and \(f_{22}=12\%\), compared to \(f_{21}=91\%\) and \(f_{22}\simeq3\%\) for the full halo sample. Similarly, for \(x_{\mathrm{HI}}^{\rm V}=0.3\) and \(x_{\hi}=0.8\), we find \(f_{21}=68\%\) and \(f_{22}=11.4\%\), compared to \(60\%\) and \(2.4\%\), respectively, for the full sample. Thus, selecting more massive halos significantly increases the predicted fraction of strong absorbers, although the resulting \(f_{22}\) values remain below those inferred by \citet{Mason2025}.
Neither \citet{Heintz2025} nor \citet{Mason2025} report the stellar masses for their galaxies. However, \citet{Pollock2026} estimate a minimum halo mass of $\sim10^{9.76},M_\odot$ and a median halo mass of $\sim10^{10.75}~M_\odot$ for their sample. Therefore, the relatively low halo masses probed by our simulations could partly explain the lack of sightlines with high values of $f_{22}$ in Case-III even when we have fully neutral gas withing the virial radius. We further investigate the dependence of the \lya\ damping-wing absorption on halo and stellar mass in Section~\ref{sec:mass-HI-relation}, where we use all three simulation boxes.

Finally, we note two limitations of the present analysis. First, \citet{Behera2026} argued that reproducing the observed relation between reddening inferred from the continuum and nebular emission lines requires an additional sub-grid reddening component associated with stellar birth clouds.  Here, we do not include the H~{\sc i} column density contributed by the birth clouds. Incorporating this contribution would increase the total \NHI\ and, consequently, strengthen the predicted \lya\ absorption.

Secondly, throughout this analysis we estimate the H~{\sc i} optical depth along sightlines passing through the centre of mass of each halo. In practice, however, the effective \NHI\ inferred from an observed galaxy spectrum is determined by integrating the light from individual stars, each attenuated by the H~{\sc i} column density along its own line of sight. We find that the H~{\sc i} optical depth measured towards the halo centre is systematically higher than this luminosity-weighted effective H~{\sc i} column density. Consequently, our approach is likely to overestimate the contribution of circumgalactic H~{\sc i} to the observed \lya\ damping-wing absorption.

The results presented in this section clearly demonstrate that reproducing the observed statistics of \lya\ absorption profiles in galaxies at $z\gtrsim8$ requires an accurate determination of the neutral hydrogen fraction not only in the IGM but also in the gas within and surrounding galaxies. This underscores the need for self-consistent modelling of the ionization state of the circumgalactic environment, in addition to that of the large-scale IGM, when interpreting \lya\ damping-wing observations during the EoR.

}

\section{Discussion}
\label{sec:discussion}
\subsection{Comparing the galaxy-\lya\ profiles among the boxes}
\label{subsec:comparison}

We compare the Ly$\alpha$ absorption profiles of galaxy sightlines among the three simulation boxes using halos in the common mass range of $10^{10.5}$--$10^{11.5}~M_\odot$ (same sample used for  Figure~\ref{fig:rad_profile2}) at $z=8$ for Case-II. As seen previously, the largest differences in the radial gas over-density occur within $\lesssim2$~pkpc of the galaxies, with the L50N1008 simulation exhibiting the highest central over-density. Here, we examine whether these differences produce measurable variations in the Ly$\alpha$ absorption properties. The values of $f_{21}$, $f_{\rm DLA}$, and $\langle N_{\hi}\rangle$ are listed in Table~\ref{table:xp_3box}.

The three simulations exhibit the same qualitative dependence of the Ly$\alpha$ absorption on $R_{\hii}$ and \xhi\ as found for the full galaxy sample. The incidence of strong absorbers ($N_{\hi}>10^{21}~\mathrm{cm}^{-2}$) decreases with increasing $R_{\hii}$ and increases with increasing \xhi.
The two $50~h^{-1}$~cMpc simulations predict nearly identical values of $f_{21}$ and $f_{\rm DLA}$ for all combinations of $R_{\hii}$ and \xhi. However, the $\langle N_{\hi}\rangle$ values for $R_{\hii}=0~\mathrm{pkpc}$ in higher-resolution L50N1008 simulation are systematically larger by $\sim0.1$~dex. Since we found good convergence for IGM sightlines (see Table~\ref{table:xp_z_IGM}), this likely reflects incomplete numerical convergence of the column density contribution from neutral gas in ISM of the halos which has the most significant contribution in case of $R_{\hii}=0~\mathrm{pkpc}$. Nevertheless, the offset is substantially smaller than the typical uncertainties in H~{\sc i} column densities inferred from JWST observations. 

We next compare the L50N1008 and L140N1008 simulations.
In L140N1008 box, at $z\sim8$, $R_{\hii}\le200~\mathrm{pkpc}$ for \xhi\ = 1.0 is able to produce $f_{21}$ within the observed range reported in \citep{Heintz2025}. 
However, $f_{22}$ values for $R_{\hii}\ge50~\mathrm{pkpc}$ are $<1\%$ even in a completely neutral medium. 
At $R_{\hii}=0$~pkpc, both simulations predict similar values of $f_{21}$, although the mean column densities in L140N1008 are lower by $\sim0.22$--0.26 dex for the three values of \xhi. Since L140N1008 has a lower mass resolution than L50N1008, this behavior is consistent with the resolution dependence discussed above. {For models with $R_{\rm HII}\ge 50$~pkpc and \xhi\ $\ge0.8$, however, L140N1008 predicts slightly stronger absorption than L50N1008, with $f_{21}$ larger by  $\le14\%$ and $\langle N_{\hi}\rangle$ higher by $\le0.08$~dex.} This trend is consistent with the conclusion for the IGM sightlines (Case~I) where for a given \xhi, the $f_{21}$ in L140N1008 is $\sim 6\%$ higher than L50N1008 at $z\sim 8$ (see Table~\ref{table:xp_z_IGM}).

\subsection{Influence of halo mass on the DLA profiles}
\label{sec:mass-HI-relation}
\begin{figure*}
\begin{minipage}{\textwidth}
\centering
\includegraphics[width=1.\textwidth]{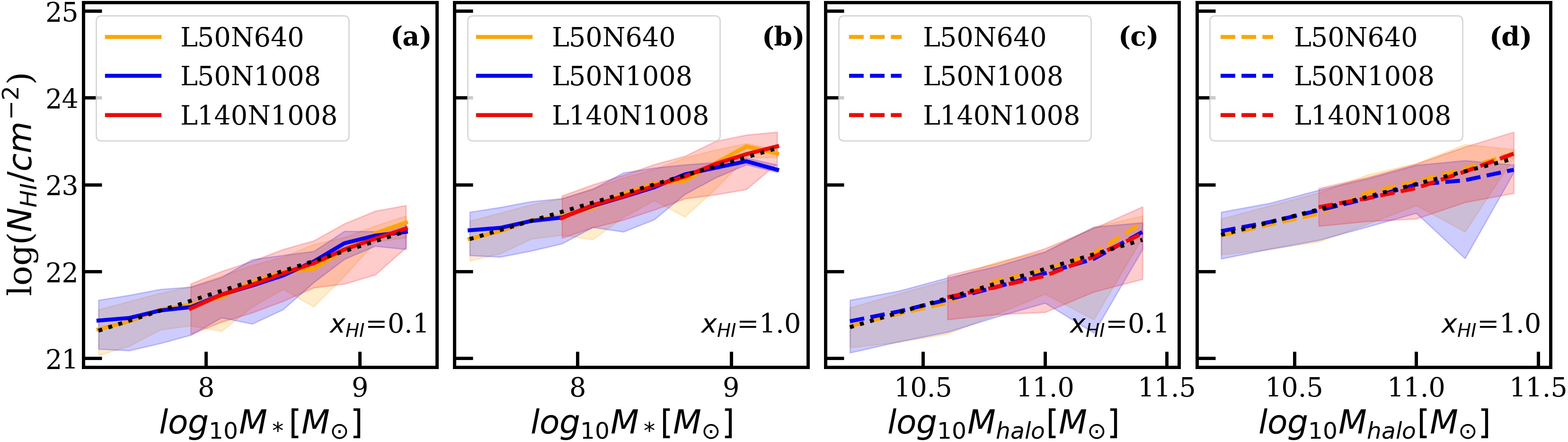}
	\caption{The median H~{\sc i} column densities along the galaxy sightlines for $R_{\rm HII}=0$~pkpc are plotted in logarithmic bins of $0.2$ dex, with the stellar mass (in the first two panels on the left) and the halo mass (in two panels on the right) of the identified halos. The 1st and 3rd panels show results for \xhi\ $=0.1$, and, 2nd and 4th panels show results for \xhi\ $=1.0$, respectively. The shaded regions show $1\sigma$ range of the $N_{\hi}$ values within the bins. The column densities obtained from L50N640 and L140N1008 are rescaled with respect to L50N1008. The dotted black lines show the best fit parameters for a power law relation between median $N_{\hi}$ for a combined sample from all three boxes and stellar or halo mass.	}
\label{fig:NHI_Mhalo_Mstar}
\end{minipage}%
\end{figure*}

As discussed earlier, cosmological simulations predict a correlation between halo (or stellar) mass and \NHI\
\citep[e.g.,][]{Gelli2025,Steen2026}. In Figure~\ref{fig:profiles}, the \(x_p\) values at $R_{\mathrm{HII}}=0$~pkpc show substantial scatter for both \xhi ~=0.1 and 1.0, a trend seen in all three simulation boxes. Since this scatter is most pronounced for \(R_{\hii}=0\) pkpc, it suggests that the strength of the \lya\ absorption depends on the intrinsic properties of the host halo, even in the absence of a local ionized bubble. We therefore investigate how the H~\textsc{i} column density inferred from the simulated \lya\ absorption depends on halo and stellar mass for the Case-II model at $z\sim8$, considering $R_{\mathrm{HII}}$=0~pkpc and \xhi ~=0.1 and 1.

Figure~\ref{fig:NHI_Mhalo_Mstar} shows the median column density of the \lya\ absorbers in 0.2 dex stellar and halo mass bins, with shaded regions indicating the 1$\sigma$ range in the $N_{\hi}$ distribution. As discussed in Section~\ref{subsec:comparison}, for a common stellar or halo mass range, the L50N640 and L140N1008 simulations yield systematically lower column densities than L50N1008. Since L50N1008 has the highest particle resolution, the sightlines in this box are expected to trace the neutral hydrogen in the vicinity of the galaxies most accurately. Therefore we rescale the column densities of the other two simulations by constant multiplicative factors of 1.19 and 2.31, respectively. The corrected median column densities are shown by the yellow and red curves, together with the L50N1008 results in blue. We then combine the column density values from all three simulations spanning a wider mass range and find the best fit parameters for a power-law relation between the median column density of the whole sample and the stellar/halo mass bins. We derive the values of the parameters (A, B) corresponding to the relation $N_{\hi} = 10^A \left(\frac{M_{\mathrm{stellar~or~halo}}}{10^{11}M_\odot} \right )^B $ for each box.

We find that the median $N_{\hi}$ values increase with increasing stellar mass of the halos. There is a strong correlation between stellar mass and median values of $N_{\hi}$, given by the correlation coefficient of 0.66 and p-value $<0.05$. The value of correlation coefficient is similar for both values of \xhi. The values of the slope (B parameter) for \xhi\ $=0.1$ and 1.0 are 0.57$\pm$0.02 and 0.52$\pm$0.018, respectively. The median $N_{\hi}$ values also positive correlation with the halo mass, as indicated by the correlation coefficient of $\sim 0.60$ and p-value of $<0.05$. The slope of the median $N_{\hi}$ and halo mass relation is steeper than that with the stellar mass. The values of B parameters for $N_{\hi}-M_{\mathrm{halo}}$ relation are 0.84$\pm$0.06 and 0.73$\pm$0.03 for \xhi\ $=0.1$ and 1.0, respectively. The scaling of DLA column densities with the stellar and halo mass of the halos implies that most of the massive halos in the simulations reside in denser environments, resulting in stronger \lya\ absorption. Similar correlations between \NHI\ and halo mass have been reported by \citet{Gelli2025} and \citet{Steen2026} using the {\sc SERRA} and {\sc TECHNICOLOR DAWN} simulations. However, since the methods used to estimate $N_{\hi}$ differ, we do not compare the fitted slopes or intercepts directly.

\section{Summary and conclusion}
\label{Sec:conclusion}

We investigate the origin of the \lya\ damping wing observed in galaxies at \(z\sim8\)–10 using large-scale cosmological hydrodynamical simulations from the {\sc Ninja} suite \citep[see][for details]{Behera2026}. Our analysis is based on three simulations spanning different volumes and mass resolutions (table~\ref{tab:sim_details}): L50N640 and L50N1008 (\(50~h^{-1}\) Mpc, containing \(2\times640^3\) and \(2\times1008^3\) particles, respectively) and L140N1008 (\(140~h^{-1}\) Mpc, \(2\times1008^3\) particles). 
At \(z\sim8\), the two \(50~h^{-1}\) Mpc simulations predominantly sample halos with masses of \(10^{9.5}\)–\(10^{10.5}~{\rm M_\odot}\), corresponding to a median stellar mass of \(\sim10^{7.5}~{\rm M_\odot}\), whereas 98\% of the halos in L140N1008 lie in the mass range \(10^{10.5}\)–\(10^{11.5}~{\rm M_\odot}\), with a median stellar mass of \(\sim10^{8.2}~{\rm M_\odot}\) (Table~\ref{tab:M_star_list}). 
This combination allows us to investigate the origin of the \lya\ damping wing over nearly two orders of magnitude in halo mass while simultaneously assessing the numerical convergence of our results.

To isolate the contributions of the diffuse IGM, the clustered gas surrounding galaxies, and neutral gas within the virial radius to the observed \lya\ damping wing, we construct mock \lya\ spectra using three idealized models. In Case~I, the \lya\ damping wing arises solely from a diffuse IGM with a uniform neutral fraction \(x_{\hi}\). Case~II extends this model by placing each galaxy inside an H~\textsc{ii} region of radius \(R_{\hii}\), thereby accounting for the clustered gas surrounding galaxies while assuming no neutral gas within the H~\textsc{ii} region. Case~III further includes neutral gas within the halo virial radius by assigning it a residual neutral fraction \(x_{\rm HI}^{\rm V}\), while the gas between the virial radius and \(R_{\hii}\) remains ionized. We explore models with \(x_{\hi}=0.1\), 0.8, and 1.0, and H~\textsc{ii} region radii spanning \(0\le R_{\hii}\le400\) pkpc. The resulting Ly\(\alpha\) absorption is characterized using four statistics: \(f_{\rm DLA}\), the fraction of sightlines with \(N_{\hi}>10^{20.3}~\mathrm{cm^{-2}}\); \(f_{21}\) and \(f_{22}\), the fractions with \(N_{\hi}>10^{21}\) and \(10^{22}~\mathrm{cm^{-2}}\), respectively; and the median \NHI, \(\langle N_{\hi}\rangle\). We compare these predictions primarily with the observational measurements of \citet{Heintz2025} and \citet{Mason2025}. Our main findings are summarized below.

\begin{enumerate}

\item
The median gas over-density along sightlines decreases with distance from the halo centre and approaches the mean IGM value at a characteristic radius, \(R_{\rm eq}\). Considering all halos in each simulation, we find \(R_{\rm eq}\simeq66\) pkpc for the two \(50~h^{-1}\) Mpc boxes, whereas \(R_{\rm eq}\simeq105\) pkpc for the more massive halos in L140N1008, showing that the extent of the overdense environment increases systematically with halo mass. To assess numerical convergence, we restrict the analysis to halos in the common mass range \(10^{10.5}\)–\(10^{11.5}~{\rm M_\odot}\). For this subsample, all three simulations yield \(R_{\rm eq}\simeq100\) pkpc, consistent within the quoted scatter. We find little evolution in the radial over-density profiles between \(z\sim8\) and \(z\sim10\), indicating that most of the clustered-gas contribution to the \lya\ damping wing originates within \(\sim100\) pkpc of galaxies, with its extent determined primarily by halo mass rather than redshift over the interval considered here.

\item
At both \(z=8\) and \(z=10\), none of the Case~I sightlines reach the DLA threshold for \(x_{\hi}=0.1\), whereas for \(x_{\hi}\ge0.8\), nearly all sightlines exhibit DLA absorption (\(f_{\rm DLA}\sim100\%\)). This is inconsistent with the high incidence of damping wings observed in \(z>8\) JWST galaxies, implying that the mean \xhi\ of the IGM must be substantially higher than 0.1 if the \lya\ absorption arises solely from a uniformly ionized IGM. The fraction of stronger absorbers, \(f_{21}\), also increases rapidly with both the \xhi\ and $z$. At \(z\sim8\), \(f_{21}\) rises from 38\% for \(x_{\hi}=0.8\) to 81\% for \(x_{\hi}=1.0\), while at \(z\sim10\) it is already \(\sim97\%\) for \(x_{\hi}=0.8\) and reaches 100\% for a fully neutral IGM. The results are well converged across the simulations, with differences of less than 1\% at \(z=10\) and at most 6\% at \(z=8\). However, even for a fully neutral IGM, none of the simulated sightlines attain \(N_{\hi}>10^{22}~\mathrm{cm^{-2}}\). Since a significant fraction of \(z>8\) galaxies exhibit such strong absorbers \citep{Mason2025,Heintz2025}, a diffuse IGM alone cannot explain the observations.

\item
Compared to random IGM sightlines (Case~I), galaxy-centred sightlines (Case~II) produce systematically stronger Ly\(\alpha\) damping-wing absorption (\(f_{21}\), \(f_{22}\), and \(\langle N_{\hi}\rangle\)) for \(R_{\hii}\lesssim100\) pkpc, comparable to the characteristic extent of the overdense gas surrounding galaxies. As expected for larger H~\textsc{ii} regions, the absorption becomes weaker than in Case~I. In the absence of local ionized regions (\(R_{\rm HII}=0\)), more than 90\% of galaxies have \(N_{\hi}>10^{21}~\mathrm{cm^{-2}}\), substantially exceeding the observed incidence, implying that a good fraction of galaxies must be surrounded by finite-sized ionized regions. At \(z\sim8\), the observed \(f_{21}\) is reproduced by models with \(R_{\hii}\lesssim200\) pkpc for \(x_{\hi}=1\) and \(R_{\hii}\lesssim100\) pkpc for \(x_{\hi}=0.8\). At \(z\sim10\), the corresponding limits increase to \(\lesssim400\) and \(\lesssim200\) pkpc, respectively. However, once \(R_{\hii}\gtrsim50\) pkpc, the predicted fraction of the strongest absorbers (\(f_{22}\)) falls below 1\%, even for a fully neutral IGM. Thus, while Case~II successfully reproduces the observed \(f_{21}\), it cannot account for the observed incidence of the strongest absorbers.

\item
In Case~III, we include Ly\(\alpha\) absorption from residual neutral gas within the virial radius while assuming that the gas between the virial radius and \(R_{\hii}\) remains fully ionized. Including this component substantially increases the incidence of the strongest absorbers, particularly in massive halos. For \(x_{\rm HI}^{V}=x_{\rm HI}=1\) and \(R_{\hii}=50\) pkpc, \(f_{22}\) increases from \(\lesssim1\%\) in Case~II to 3\% for the full halo sample and to 12\% for halos with masses \(10^{10.5}\)–\(10^{11.5}\,{\rm M_\odot}\). Residual neutral gas within the virial radius is therefore a necessary ingredient for producing the strongest damped Ly\(\alpha\) absorbers. However, for the halo masses resolved in our simulations, it is not sufficient to reproduce the observed incidence. This suggests that the observed galaxies preferentially reside in more massive halos and/or that additional unresolved neutral gas, such as dense stellar birth clouds, contributes to the absorption.

We also note that our synthetic spectra are constructed using sightlines originating from the halo centres. In reality, the observed spectra probe individual star-forming regions distributed throughout galaxies, and the corresponding effective \NHI\ is likely to be lower on average. A more realistic treatment of the source locations would therefore reduce the predicted incidence of the strongest absorbers, making the discrepancy with the observations even larger and further strengthening the need for an additional unresolved neutral component.

\item
To isolate the dependence of Ly\(\alpha\) absorption on galaxy properties, we analysed the Case~II models with \(R_{\hii}=0\). We find that the inferred \NHI\ along galaxy sightlines increases with both stellar and halo mass, reflecting the larger gas reservoirs and denser environments of more massive halos. Consequently, the incidence of the strongest damped Ly\(\alpha\) absorbers is expected to depend sensitively on the halo-mass distribution of the observed galaxy sample. Future measurements of the stellar and halo masses of JWST galaxies will provide a direct test of this prediction and enable more robust comparisons between simulations and the observed incidence of the strongest damped Ly\(\alpha\) absorbers.

\end{enumerate}

Despite differences in simulation volume and mass resolution, the statistical properties of the predicted \lya\ damping-wing absorbers are broadly consistent across our three simulation boxes, indicating that our main conclusions are numerically robust over the halo-mass range resolved by our simulations. The observed \lya\ damping-wing statistics at \(z\gtrsim8\) cannot be explained solely by the neutral fraction of the diffuse IGM, but also depend sensitively on the distribution of H~{\sc i} within galaxies and their surrounding circumgalactic environment. Accurately interpreting these observations therefore requires modelling the ISM, CGM, and IGM within a unified radiative-transfer framework.

Current theoretical approaches generally emphasize one of two complementary aspects of this problem. High-resolution radiative-transfer simulations can model the escape of ionizing photons from the ISM and the resulting structure of the circumgalactic medium in considerable detail, but are typically limited to a small number of galaxies (e.g. \citet{Gelli2025}) or relatively small cosmological volumes (e.g., \citet{Steen2026}). For example, \citet{Gelli2025} analyzes only $\sim100$ galaxies at $z=6\text{--}9.5$ from the SERRA simulations, a high-resolution cosmological zoom-in simulation with on-the-fly RT, while \citet{Steen2026} uses the TECHNICOLOR DAWN simulation with a box side length of $16.5~h^{-1}~\mathrm{cMpc}$, limiting the number of the most massive halos formed in the simulation. Conversely, large-scale reionization simulations capture the growth of ionized regions and the collective contribution of galaxies over cosmological volumes, including the rare massive halos that are expected to host the brightest JWST galaxies, but generally do not resolve the small-scale ISM physics that regulates the escape of ionizing radiation \textcolor{black}{(see e.g., \citet{Keating2023,Keating2024})}. Our results demonstrate that both aspects are essential for interpreting \lya\ damping-wing observations, highlighting the need to connect small-scale radiative-transfer calculations around galaxies with large-scale reionization simulations in a self-consistent manner.

While our treatment of reionization is intentionally idealized and does not include self-consistent RT, it captures the key physical trends needed to interpret the low-resolution JWST spectra. Future work combining cosmological hydrodynamical simulations with multi-scale radiative-transfer calculations spanning the ISM, CGM, and IGM, together with improved constraints on the emergent \lya\ line profiles of high-redshift galaxies, will enable more stringent tests of these conclusions.

\section*{Acknowledgments}
{
We acknowledge the use of the high-performance computing facilities PERSEUS and PEGASUS at IUCAA and SAHA, CHANDRA and KALINGA at NISER for numerical work presented here. We would like to thank Simeon Bird for useful clarifications and discussions regarding MP-GADGET and for fixing bugs that came up during this simulation campaign. NK acknowledges support from the IUCAA Associateship Programme.}

\section*{Data Availability}
The data generated in this article are available from the corresponding author.



\bibliographystyle{aa} 
\bibliography{main} 

\appendix

\section{Galactic emission and ionizing bubble radius}
\label{sec:galactic_emission}
In this work, we have assumed a fixed value of \(R_{\hii}\) in our model. In reality, however, the sizes of ionized bubbles are expected to vary from one halo to another, depending on several factors, including the fraction of ionizing radiation that escapes from the galaxy, the density of the surrounding intergalactic medium, and the contribution from neighboring ionizing sources. In this section, we estimate a representative value of \(R_{\hii}\) for one of the brightest galaxies in our simulations.

To this end, we identify the star particles associated with one of the most massive halos in the L50N1008 simulation and reconstruct its star formation history. The intrinsic stellar emission is modelled using the spectral energy distribution of \citet{bruzual2003}, adopting the Padova 1994 evolutionary tracks and a Chabrier initial mass function \citep{chabrier2003}. For simplicity, we assume a constant stellar metallicity equal to the solar value throughout the galaxy's evolution. For a halo with a stellar mass of \(10^{8.1}~{\rm M_\odot}\) at \(z=10\), we estimate a hydrogen-ionizing \((\lambda < 912,\mathrm{\AA})\) photon production rate of approximately \(\Ndot\simeq10^{53.4}\) photons s\(^{-1}\).

A fraction of these ionizing photons escapes the dense gas surrounding the galaxy and ionizes the ambient IGM, producing an expanding H~\textsc{ii} region. As the ionization front propagates from \(R_{\hii}\) to \(R_{\hii}+dR_{\hii}\), the number of newly ionized hydrogen atoms is \(4\pi R_{\hii}^{2}n_{\rm H}(R_{\hii})~dR_{\hii}\), while the total recombination rate within the ionized bubble is given by \(\int 4\pi R_{\hii}^{2}\alpha_{\hi}n_i(R_{\hii})n_e(R_{\hii}),dR_{\hii}\). The evolution of the ionization front is therefore governed by

\begin{equation}
    \frac{d R_{\hii}}{dt}= \frac{\Ndot -\int 4 \pi R_{\hii}^2 \alpha_{\hi} n_{i}(R_{\hii}) n_{e}(R_{\hii}) dR_{\hii}}{4 \pi R^2_{\hii} x_{\hi} n_H(R_{\hii})} 
\end{equation}
where $\Ndot$  is the rate of hydrogen-ionizing photons escaping into the IGM, \(\alpha_{\hi}\) is the hydrogen recombination coefficient, \(n_i\), \(n_e\), and \(n_{\rm H}\) are the number densities of ions, electrons, and hydrogen, respectively, and \xhi is the neutral hydrogen fraction. We assume that the surrounding medium is initially fully neutral \((x_{\hi}=1)\).

Using the median gas density profile around the selected halo and assuming an escape fraction of unity, we obtain a characteristic H~\textsc{ii} region radius of \(\sim400\) pkpc. This estimate should be regarded as an upper limit
as the escape fraction is expected to be less than 1. Reionization models generally favour escape fractions of \(f_{\rm esc}\sim0.2\) at \(z\sim8\) \citep{Mitra2013, robertson2015, khaire2016}, although \(f_{\rm esc}\) is expected to depend on halo mass \citep{chakraborty2024}. Recent JWST/NIRSpec observations of three galaxies at \(z\sim7.5\) by \citet{Jung2024} further illustrate this diversity. They infer \(f_{\rm esc}\simeq0.64\) for one galaxy, capable of producing an ionized bubble of nearly 1 pMpc within 30--40 Myr of star formation, while the other two galaxies, with \(f_{\rm esc}\simeq0.08\) and \(<0.06\), are expected to produce H~\textsc{ii} regions of approximately 200 and 400 pkpc, respectively, over a \(\sim50\) Myr star formation episode.

Although the bubble size around an individual galaxy can span a wide range depending on its physical properties and environment, these simple estimates indicate that ionized regions with radii of a few hundred physical kiloparsecs are expected for luminous galaxies at \(z\sim10\). We therefore conclude that the range of \(R_{\hii}\) adopted in our toy models is physically well motivated and broadly consistent with both theoretical expectations and current observational constraints.

\section{Details of the simulations}
Table~\ref{tab:sim_details} lists the boxsize, 
\begin{table*}[]
    \centering
    \caption{Details of the simulations used in this study. The columns denote the run LXNY with X and Y representing $L_{\rm box}$ (column 2, the comoving box size) and $N_{\rm part}$ (column 3, the initial number of particles dark matter and gas), the logarithm of mass of the dark matter $m_{\rm DM}$ , initial gas $m_{\rm gas}$ and star $m_{\star}$ particles,  in units of solar masses $M_{\odot}/h$. The mass of the star particle is roughly a fourth of the initial mass of the gas particle.}
    \begin{tabular}{lccccc}
         &  \\
    \hline
    Run ID 
    & $L_{\rm box}$ 
    & $N_{\rm part}$   
    & $\log \left(m_{\rm DM}/h^{-1}M_{\odot}\right)$ 
    & $\log\left(m_{\rm gas}/h^{-1}M_{\odot}\right)$ 
    & $\log\left(m_{\star}/h^{-1}M_\odot\right)$\\
        & $(h^{-1} \mathrm{Mpc})$  & & & &  \\  
    \hline
    \hline
         L140N1008 &140 & $2\times 1008^3$ & 8.24 & 7.54 & 6.94 \\
         L50N1008 & 50  & $2\times 1008^3$ & 6.90 & 6.20 & 5.59 \\
         L50N640  & 50  & $2\times 640^3 $  & 7.49 & 6.79 & 6.19 \\
         \hline         
    \end{tabular}
    \label{tab:sim_details}
\end{table*}

\section{Properties of the halo samples in the three simulations}
The number of halos and median stellar mass of the sample of halos from the three simulations used in this work are summarized in Table~\ref{tab:M_star_list}.
\begin{table*}
    \centering  
   \caption{Properties of the halo samples identified in the three simulations used in this study. For each simulation, we list the number of halos ($N_h$) identified with {\sc Rockstar-Galaxies} and the logarithm of the median stellar mass, log$(M_*/M_\odot)$.}
   \begin{threeparttable}
    \begin{tabular}{c|cc|cc|cc}
    \hline
     Redshift &    \multicolumn{2}{c|}{L140N1008} & \multicolumn{2}{c|}{L50N1008} & \multicolumn{2}{c}{L50N640} \\
     &$N_{h}$  &$\log \left(\frac{M_{*}}{M_\odot}\right)$  &$N_{h}$  &$\log \left(\frac{M_{*}}{M_\odot}\right)$  &$N_{h}$  &$\log \left(\frac{M_{*}}{M_\odot}\right)$  \\
    \hline 
    \hline
    z=8   & 1216 & 8.24 & 772 & 7.44 & 555 & 7.54\\
    z=10   & 59 & 8.22 & 82 & 7.39 & 54 & 7.51\\
    \hline
    \end{tabular}
    \label{tab:M_star_list}
     \end{threeparttable}    
\end{table*}

\section{Redshift evolution of the radial overdensity profiles}
\begin{figure*}
\begin{minipage}{\textwidth}
\centering
\includegraphics[width=0.8\textwidth]{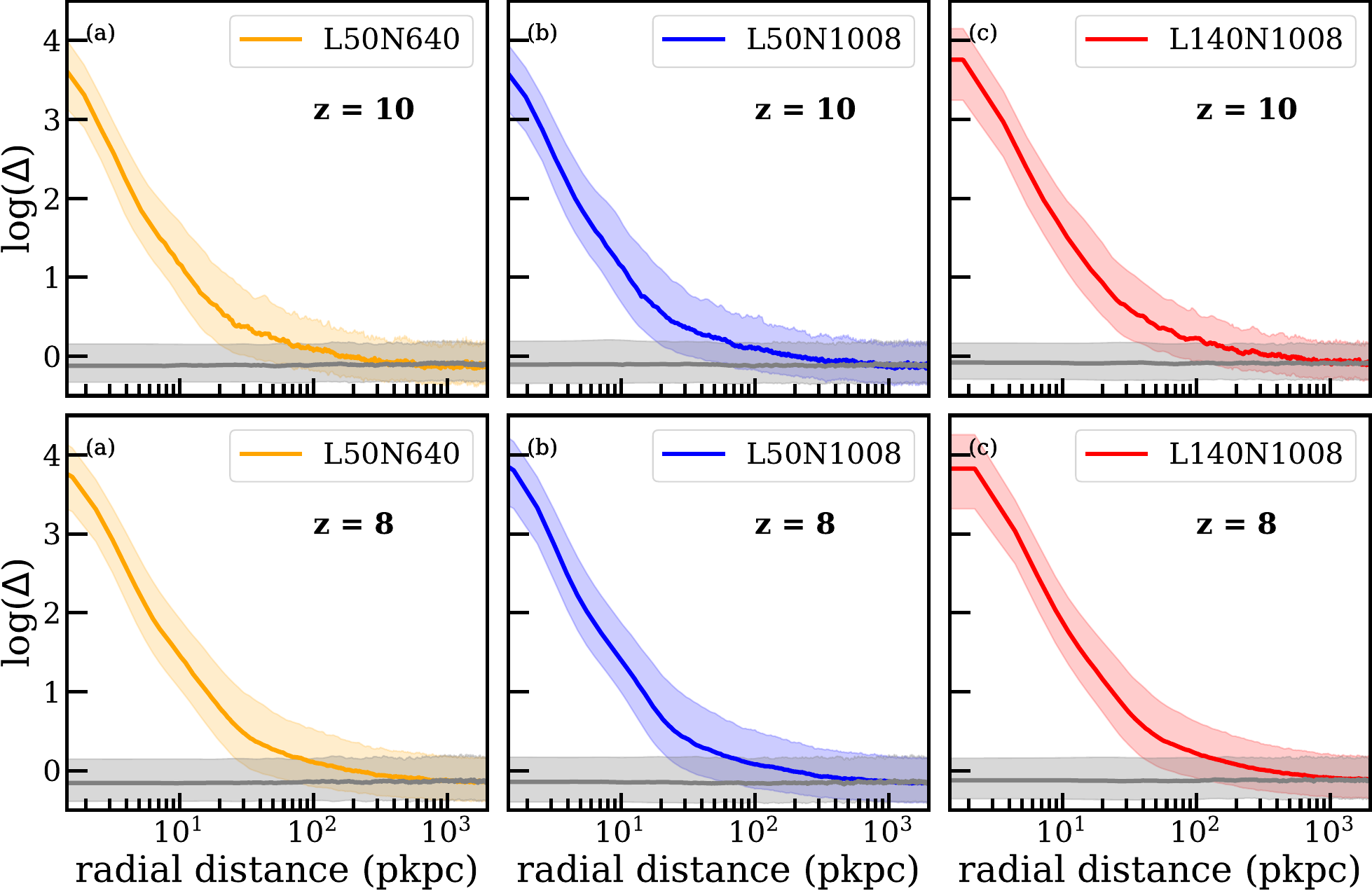}
	\caption{The median over-density profile along the radial distance for sightlines drawn around the halos in L50N640, L50N1008 and L140N1008 boxes are shown as orange, blue and red colored lines in the left, middle, and right panels, respectively, for redshift 10 and 8 in the upper and lower panels. The shaded region shows the 1$\sigma$ range of over-densities for the different sightlines along all the identified halos in the simulation boxes. The radial over-density profile along the IGM sightlines are shown in grey lines. The over-density in all three simulations becomes comparable to the IGM over-density profiles around 100pkpc for both redshift 10 and 8. 
	}
\label{fig:rad_profile1}
\end{minipage}%
\end{figure*}

Figure~\ref{fig:rad_profile1} shows the radial profiles of the median overdensities along the sightlines considered in Case II for $z=8$ and 10 in the upper and lower panels, respectively.  

\section{Summery of parameters characterizing the absorbers in the sightlines} The $f_{21}$, $f_{\rm DLA}$, log $\langle N_{\hi}\rangle$ values for the IGM sightlines considered in Case~I are listed in Tables~\ref{table:xp_z_IGM}. The \lya\ absorption profiles along the galaxy sightlines (considered in Case II) for varied $R_{\mathrm{HII}}$ in L50N1008 box are shown in Figure~\ref{fig:profiles}. Table~\ref{table:xp_z8_gal} and Table~\ref{table:xp_z10_gal} summarizes these parameters for the galaxy sightlines (Case~II) at $z=8$ and 10, respectively. 

\begingroup
\setlength{\tabcolsep}{4pt} 
\begin{table*}
    \centering
    \caption{Different parameters characterizing the absorbers in the galaxy sightlines (case~II) at z=10 for the two 50$h^{-1}$cMpc boxes. The description of $f_{21}$, $f_{\rm DLA}$ and log $\langle N_{\hi}\rangle$ are same as Table~\ref{table:xp_z_IGM}.}
     \begin{threeparttable}
   \begin{tabular}{c|ccc|ccc|ccc|ccc|ccc}
   \hline
   & \multicolumn{3}{c|}{$R_{\rm HII}=0$ pkpc} & \multicolumn{3}{c|}{$R_{\rm HII}=50$pkpc} & \multicolumn{3}{c|}{$R_{\rm HII}=100$ pkpc} & \multicolumn{3}{c|}{$R_{\rm HII}=200$pkpc} & \multicolumn{3}{c}{$R_{\rm HII}=400$ pkpc} \\

     \xhi & $f_{21}$  & $f_{\rm DLA}$  & log~$\langle N_{\hi}\rangle $  & $f_{21}$ & $f_{\rm DLA}$ & log~$\langle N_{\hi}\rangle $  & $f_{21}$ & $f_{\rm DLA}$ &  log~$\langle N_{\hi}\rangle $  & $f_{21}$ & $f_{\rm DLA}$ &  log~$\langle N_{\hi}\rangle $  & $f_{21}$ & $f_{\rm DLA}$ & log~$\langle N_{\hi}\rangle $   \\
     \hline
\multicolumn{16}{c}{For L50N1008 at z=10} \\
0.1 & 94 & 100 & 21.58 & 0 & 3 & 19.33 & 0 & 1 & 19.01 & 0 & 0 & - & 0 & 0 & -\\
0.8 & 100 & 100 & 22.51 & 99 & 100 & 21.25 & 89 & 100 & 21.16 & 52 & 100 & 21.01 & 5 & 96 & 20.66 \\
1.0 & 100 & 100 & 22.62 & 100 & 100 & 21.40 & 100 & 100 & 21.32 & 92 & 100 & 21.19 & 30 & 100 & 20.91\\
\multicolumn{16}{c}{For L50N640 at z=10} \\
0.1 & 97 & 100 & 21.59 & 0 & 3 & 19.24 & 0 & $\sim$1 & 18.99 & 0 & 0 & - & 0 & 0 & -\\
0.8 & 100 & 100 & 22.52 & 98 & 100 & 21.24 & 86 & 100 & 21.15 & 47 & 100 & 20.99 & 3 & 96 & 20.66\\
1.0 & 100 & 100 & 22.62 & 100 & 100 & 21.39 & 100 & 100 & 21.31 & 90 & 100 & 21.17 & 27 & 100 & 20.90\\

\hline
    \end{tabular}
    \label{table:xp_z10_gal}

     \end{threeparttable}
\end{table*}
\endgroup

\section{Summary of the comparison of the galaxy-\lya\ profiles among the boxes} Table~\ref{table:xp_3box} summarizes the comparison of $f_{21}$, $f_{\rm DLA}$, log $\langle N_{\hi}\rangle$ values obtained from the galaxy sightlines among the three simulation boxes. We considered sightlines around the halos within the mass range of $10^{10.5}-10^{11.5}\mathrm{M}_{\odot}$ for this comparison. 
\begingroup
\setlength{\tabcolsep}{5pt} 
\begin{table*}
    \centering
    \caption{Comparison of absorption in the galaxy sightlines (case~II) among the L50N1008, L50N640 and L140N1008 boxes within halo mass range $10^{10.5} - 10^{11.5}M_\odot$. The description of $f_{21}$, $f_{\rm DLA}$ and log $\langle N_{\hi}\rangle$ are same as Table~\ref{table:xp_z_IGM}.}
     \begin{threeparttable}
   \begin{tabular}{c|ccc|ccc|ccc|ccc|ccc}
   \hline
   & \multicolumn{3}{c|}{$R_{\rm HII}=0$ pkpc} & \multicolumn{3}{c|}{$R_{\rm HII}=50$ pkpc} & \multicolumn{3}{c|}{$R_{\rm HII}=100$ pkpc} & \multicolumn{3}{c|}{$R_{\rm HII}=200$ pkpc} & \multicolumn{3}{c}{$R_{\rm HII}=400$ pkpc}\\

     \xhi & $f_{21}$  & $f_{\rm DLA}$  & log~$\langle N_{\hi}\rangle $  & $f_{21}$ & $f_{\rm DLA}$ & log~$\langle N_{\hi}\rangle $  & $f_{21}$ & $f_{\rm DLA}$ & log~$\langle N_{\hi}\rangle $  & $f_{21}$ & $f_{\rm DLA}$ & log~$\langle N_{\hi}\rangle $  & $f_{21}$ & $f_{\rm DLA}$ & log~$\langle N_{\hi}\rangle $  \\
     \hline
     
\multicolumn{16}{c}{For L50N1008 } \\
0.1 & 95 & 100 & 21.75 & <1 & 4 & 19.16 & <1 & 1 & - & 0 & <1 & - & 0 & 0 & -  \\

0.8 & 100 & 100& 22.67 & 65 & 100 & 21.05 & 45 & 100 & 20.97 & 19 & 100 & 20.82 & 3 & 78 & 20.51 \\

1.0 & 100 & 100 & 22.77 & 93 & 100 & 21.20 & 81 & 100 & 21.13 & 51 &  100 & 21.00 & 12 & 98 & 20.74 \\

\multicolumn{16}{c}{For L50N640 } \\
0.1 & 95 & 100 & 21.66 & 0 & 4 & 19.16 & 0 & 2 & - &  0 & <1 & - & 0 & 0 & —  \\

0.8 & 100 & 100 & 22.59 & 66 & 100 & 21.05 & 44 & 100 & 20.97 & 18 & 100 & 20.83 & 3 & 80 & 20.50 \\

1.0 & 100 & 100 & 22.68 & 93 & 100 & 21.21 & 81 & 100 & 21.13 & 52 & 100 & 21.01 & 12 & 98 & 20.74 \\

\multicolumn{16}{c}{For "L140N1008" } \\
0.1 & 92 & 100 & 21.49  & 0 & 3.7 & 19.32 & 0 & 1 & - & 0 & 0 & - & 0 & 0 & -\\

0.8 & 100 & 100 & 22.45 & 79 & 100 & 21.13 & 58 & 100 & 21.03 & 28 & 100 & 20.89 & 5 & 88 & 20.59\\

1.0 & 100 & 100 & 22.55 & 98 & 100 & 21.27 & 91 & 100 & 21.19 & 65 & 100 & 21.06 & 19 & 99 & 20.82 \\
\hline
    \end{tabular}
    \label{table:xp_3box}
     \end{threeparttable}
\end{table*}
\endgroup

\label{lastpage}

\end{document}